\documentclass[12pt]{article}
\usepackage[margin=20truemm]{geometry}
\usepackage[dvipdfmx]{graphicx,color}
\usepackage{bigints}
\usepackage{amsmath,amssymb,bm}
\usepackage{amsthm}
\usepackage{type1cm}
\usepackage{braket}
\usepackage{esvect}
\usepackage{cite}
\usepackage{hyperref}
\usepackage{xcolor}
\hypersetup{%
setpagesize=false,%
bookmarks=true,%
bookmarksdepth=tocdepth,%
bookmarksnumbered=true,%
colorlinks=true,%
linkcolor=black,%
citecolor=black,%
urlcolor=black,%
pdftitle={},%
pdfsubject={},%
pdfauthor={},%
pdfkeywords={}}

\allowdisplaybreaks[3]

\theoremstyle{plain}
\newtheorem{thm}{Theorem}
\newtheorem*{thm*}{Theorem}
\theoremstyle{definition}
\newtheorem{dfn}{Definition}
\numberwithin{equation}{section}
\newtheorem{lem}[thm]{Lemma}

\usepackage{tikz}
\usetikzlibrary{matrix, positioning, arrows, calc}

\begin{document}

\begin{titlepage}

\begin{flushright}   \end{flushright}

~~\\

\vspace*{0cm}
    \begin{Large}
    	\begin{center}
       {Mutation and crossover of simplicial complexes}
       \end{center}
    \end{Large}
\vspace{1cm}

\begin{center}
        Boyu L{\sc i},$^{*}$\footnote
           {e-mail address: h26ds204@hirosaki-u.ac.jp}
	Kohta H{\sc atakeyama},$^{**}$\footnote
           {e-mail address: kohta.hatakeyama@gauge.scphys.kyoto-u.ac.jp}
        Matsuo S{\sc ato},$^{*}$\footnote
           {e-mail address: msato@hirosaki-u.ac.jp}
Yuji S{\sc ugimoto}$^{***}$\footnote
           {
e-mail address: yujisugimoto@postech.ac.kr}           
 and \\
	Gota T{\sc anaka}$^{****}$\footnote
           {e-mail address: gotanak@mi.meijigakuin.ac.jp} \\
      \vspace{1cm}
         {$^{*}$\it Graduate School of Science and Technology, Hirosaki University\\
          Bunkyo-cho 3, Hirosaki, Aomori 036-8561, Japan}\\
         {$^{**}$\it Department of Physics, Kyoto University, Kyoto 606-8502, Japan}\\
 {$^{***}$\it Department of Physics, POSTECH, Pohang 37673, Korea}\\
          {$^{****}$\it Institute for Mathematical Informatics, Meiji Gakuin University,\\
1518 Kamikuratacho, Totsuka-ku, Yokohama, Kanagawa 244-8539, Japan}\\

\end{center}

\hspace{5cm}

\begin{abstract}
\noindent
Color graphs and their subgraphs, referred to as bubble graphs, correspond bijectively to the simplicial complexes of pseudomanifolds and their subsimplices, respectively. In this paper, we introduce matrix representations for colored graphs and their associated bubble graphs. By using this correspondence, we define simplicial-complex matrices and subsimplex matrices that encode the simplicial complexes of pseudomanifolds and their subsimplices. Moreover, we formulate mutation and crossover operations on colored graphs. Through the established correspondence among simplicial complexes, colored graphs, and simplicial-complex matrices, we extend these operations to simplicial complexes and simplicial-complex matrices. We further implement an algorithm generating simplicial-complex matrices and  a genetic algorithm performing  mutation and crossover of them to produce pseudomanifolds exhibiting diverse topologies. In addition, we implement procedures for decomposing the generated simplicial-complex matrices into simplex matrices, reconstructing the simplicial complexes of the associated pseudomanifolds from this information, and computing geometric quantities such as the volume, circumcenter, and dual-simplex volume of each simplex.
\end{abstract}

\vfill
\end{titlepage}
\vfil\eject

\setcounter{footnote}{0}

\section{Introduction}\label{intro}
\setcounter{equation}{0}

Superstring theory is regarded as a promising candidate for a unified theory that incorporates gravity. A central contemporary challenge in superstring theory is to determine the true vacuum among the extremely large number ($> 10^{500}$)  of perturbatively stable vacua, collectively known as the string landscape. Identifying the true vacuum would enable superstring theory to make predictions for physical phenomena consistent with experimental and observational data. One of the most reasonable approaches for determining the true vacuum is to formulate superstring theory non-perturbatively, derive the potential for string backgrounds, which represent perturbatively stable vacua, and identify its minimum. Since a string background is a solution to the equations of motion for the bosonic fields in supergravity and D-brane effective theories, the potential is a functional of these bosonic fields. Indeed, string geometry theory, which is one of the candidates of non-perturbative formulations of superstring theory \cite{Sato:2017qhj, Sato:2019cno, Sato:2020szq, Sato:2022owj, Sato:2022brv,  Sato:2023lls}, yield such a potential for string backgrounds \cite{Honda:2020sbl, Honda:2021rcd, Nagasaki:2023fnz, Sato:2025wfc, Nagasaki:2025tmi, Sato:2025qqa, futureIIB}.

Because solutions to the equations of motion are generally inaccessible analytically, it is nearly impossible to compare string backgrounds analytically through the potential itself. A viable strategy is to impose the equations of motion on the potential via the method of Lagrange multipliers, discretize the potential using Regge calculus \cite{Regge:1961px} or causal dynamical triangulations (CDT) \cite{Ambjorn:1998xu}, and then search for its minimum numerically. In particular, for the problem of minimizing the potential, no problem on the measures in the path integrals makes simple Regge calculus especially suitable.

Previous work in Regge calculus has focused primarily on simulations of a (3+1)-dimensional universe. Because the topology of the universe is believed to have been a three-sphere throughout most of its history, numerical methods have typically fixed the topology while refining the discretization. However, in string backgrounds defined in ten dimensions, the six extra dimensions beyond the observable (3+1)-dimensional spacetime carry the information relevant to elementary particle physics and our universe. Thus, it becomes necessary to determine the topology, metric, and matter fields on this six-dimensional manifold. Consequently, to locate the minimum of the potential for string backgrounds, numerical computations must be capable of generating a variety of topologies. Notably, the description of gravitational and matter fields in Regge calculus requires detailed information about the simplicial complex of the underlying manifold \cite{CHRIST1982337, Ren:1987is}.

Meanwhile, recent developments have elucidated a correspondence between the simplicial complexes of pseudomanifolds and colored graphs \cite{Gurau:2011xp}. Because colored graphs can be readily encoded as numerical data, they offer a promising route for numerically representing the simplicial complexes of manifolds. In this paper, we introduce a matrix representation of colored graphs, referred to as simplicial-complex matrices. Through this representation and the aforementioned correspondence, the simplicial complexes of manifolds can be encoded directly in matrix form. Furthermore, in genetic algorithms, mutation and crossover operations are applied to data structures in order to explore diverse configurations and solve minimization problems. Rather than defining mutation and crossover merely at the level of abstract data, we formulate these operations geometrically on colored graphs themselves. By expressing colored graphs in matrix form, we thereby define mutation and crossover operations on simplicial-complex matrices. As a result, the correspondence enables us to describe mutation and crossover of manifold simplicial complexes, allowing numerical computations to generate a broad range of topologies.

The structure of this paper is as follows. In Section 2, we review the correspondence between simplicial complexes of pseudomanifolds and colored graphs. In Section 3, we introduce the matrix representation of colored graphs and, via the correspondence from Section 2, define the matrix representation of simplicial complexes namely, simplicial-complex matrices. Section 4 presents the definitions of mutation and crossover for colored graphs and constructs their matrix representations. This in turn yields mutation and crossover operations for simplicial-complex matrices. Appendix A reviews the dual-simplex volume and the circumcenter of a simplex, which are required for discretizing matter fields. Appendix B provides the implementation of an algorithm for generating simplicial-complex matrices and  a genetic algorithm performing mutation and crossover of them. We also provide  implementations of algorithms for the numerical construction of the simplicial complexes of pseudomanifolds from the resulting matrices, as well as the computation of simplex volumes and dual-simplex volumes.

\section{A concise review of simplicial complexes and the colored-graph correspondence}

In this section, we provide a concise review of colored graphs  and their correspondence with simplicial complexes \cite{Gurau:2011xp}.

\subsection{Definitions of colored graphs and their bubbles}

\begin{figure}[h]
  \centering
  \begin{tikzpicture}[scale=1.8, thick]
    \fill (0,0) circle (0.12);
    \draw (0,0) -- (0.0, 1.2) node[above=2mm, font=\footnotesize] {0};
    \draw (0,0) -- (0.4, 1.1) node[above right=0.5mm, font=\footnotesize] {1};
    \draw (0,0) -- (0.8, 1.0) node[above right=0.5mm, font=\footnotesize] {2};
    \node at (0.6, 0.3) {\vdots};
    \draw (0,0) -- (1.1, 0.0) node[right=1mm, font=\footnotesize] {$D$};
    
    \draw (4.8,0) -- (4.8, 1.2) node[above=2mm, font=\footnotesize] {0};
    \draw (4.8,0) -- (4.4, 1.1) node[above left=0.5mm, font=\footnotesize] {1};
    \draw (4.8,0) -- (4.0, 1.0) node[above left=0.5mm, font=\footnotesize] {2};
    \node at (4.2, 0.3) {\vdots};
    \draw (4.8,0) -- (3.7, 0.0) node[left=1mm, font=\footnotesize] {$D$};

    \draw (4.8, 0) circle (0.12);
    \fill [white] (4.8, 0) circle (0.12); 
  \end{tikzpicture}
  \caption{Schematic depiction of positive (black) and negative (white) vertices in a ($D+1$)-colored graph.
Each vertex is incident to $D+1$ distinct colors, with the cyclic ordering being clockwise at positive vertices and counterclockwise at negative ones.}
  \label{Positive and negative vertex}
\end{figure}
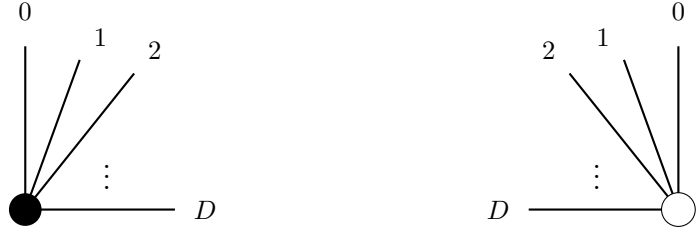

\begin{dfn}[]
A ($D+1$)-colored graph is a graph $\mathcal{G}=(\mathcal{V}, \mathcal{E})$ consisting of a vertex set $\mathcal{V}$ and an edge set $\mathcal{E}$, where each edge carries a color, and satisfying the following conditions:
\end{dfn}
\begin{itemize}
\item \textbf{Bipartition of the vertex set.}

   The vertex set $\mathcal{V}$ is partitioned into two subsets $V$ and $\bar{V}$
 of equal cardinality:
 \begin{equation}
 \mathcal{V} = V \cup \bar{V}, \qquad  2|V| = 2|\bar{V}| = |\mathcal{V}|.
\end{equation}
   Vertices in $V$ (resp. $\bar{V}$) are referred to as positive (resp. negative) vertices.
   No two vertices of the same sign are adjacent. Hence every edge $l \in \mathcal{E}$ joins a unique positive vertex $v \in V$ with a unique negative vertex $\bar{v} \in \bar{V}$, i.e., $l = \{v, \bar{v}\}$.
\item \textbf{Color partition of the edge set.}

   The edge set $\mathcal{E}$ decomposes into $(D+1)$ disjoint subsets:
 \begin{equation}
   \mathcal{E} = \bigsqcup_{i=0}^{D} \mathcal{E}_i,
 \end{equation}
   where $\mathcal{E}_i$ consists precisely of the edges of color $i$.
\item \textbf{Incidence of colored edges at each vertex.}

   Every vertex is incident to exactly $(D+1)$ edges, each of a distinct color.
   Around every positive vertex, the incident colors appear in clockwise cyclic order, whereas around every negative vertex they appear in counterclockwise order (Figure  \ref{Positive and negative vertex}).
\end{itemize}

$(D+1)$-colored graphs of this type are known to encode, via duality, $D$-dimensional pseudo-manifolds.
To make this correspondence explicit, we introduce bubbles, which are canonical subgraphs obtained by deleting edges of selected colors.
\begin{dfn}[]
For a subset of colors $\{ i_1, \dots, i_d \}$, a $(D+1-d) $-bubble is defined as a connected subgraph of $\mathcal{G}$ obtained by removing all edges whose colors belong to this set, such that every remaining vertex is incident to precisely $(D+1-d)$ edges.
\end{dfn}

A $(D+1-d)$-bubble is denoted by
$\mathcal{B}_{(\rho)}^{\hat{i}_1 \hat{i}_2 \cdots \hat{i}_d}$
where we assume $i_1 < \cdots < i_d$, and ($\rho$) specifies the set of vertices contained in the bubble.
A 0-bubble $\mathcal{B}^{\hat{1} \hat{2} \cdots \hat{D+1}}_v$ is a single vertex $v\in V$, whereas a 1-bubble
$\mathcal{B}^{\hat{1} \hat{2} \cdots \hat{k-1} \hat{k+1} \cdots \hat{D+1}}_{v\bar{v}}$
is an edge $l = \{v,\bar{v}\}$ of color $k$.

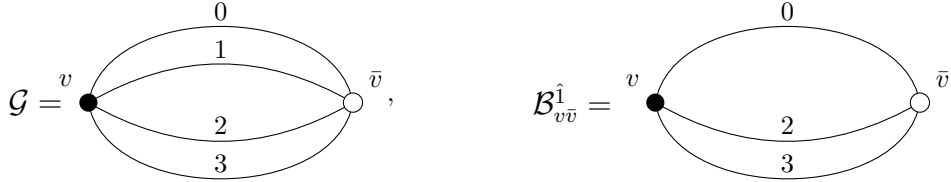
\begin{figure}[h]
  \centering
\begin{tikzpicture}
    \fill (0,0) circle (0.12);
    \draw [thick] (3.5,0) circle (0.12);

    \draw node at (-0.7,0.0) {$\mathcal{G}=$} ;
    \draw node[font=\footnotesize] at (-0.3,0.3) {$v$} ;
    \draw node[font=\footnotesize] at (3.8,0.3) {$\bar{v}$} ;
    \draw node[font=\footnotesize] at (4.0,0.0) {$,$} ;

    \draw [out=80, in=100] (0,0) to (3.5,0);
    \draw [out=30, in=150] (0,0) to (3.5,0);
    \draw [out=-30, in=-150] (0,0) to (3.5,0);
    \draw [out=-80, in=-100] (0,0) to (3.5,0);

    \draw node[font=\footnotesize] at (1.75,1.2) {0};
    \draw node[font=\footnotesize] at (1.75,0.7) {1};
    \draw node[font=\footnotesize] at (1.75,-0.3) {2};
    \draw node[font=\footnotesize] at (1.75,-0.8) {3};
    
    \fill [white] (3.5,0) circle (0.12);

    \fill (7.5,0) circle (0.12);
    \draw [thick] (11.0,0) circle (0.12);

    \draw node at (6.4,0.0) {$\mathcal{B}^{\hat{1}}_{ v \bar{v} }=$} ;
    \draw node[font=\footnotesize] at (7.2,0.3) {$v$} ;
    \draw node[font=\footnotesize] at (11.3,0.3) {$\bar{v}$} ;

    \draw [out=80, in=100] (7.5,0) to (11.0,0);
    \draw [out=-30, in=-150] (7.5,0) to (11.0,0);
    \draw [out=-80, in=-100] (7.5,0) to (11.0,0);

    \draw node[font=\footnotesize] at (9.25,1.2) {0};
    \draw node[font=\footnotesize] at (9.25,-0.3) {2};
    \draw node[font=\footnotesize] at (9.25,-0.8) {3};
    
    \fill [white] (11.0,0) circle (0.12);
\end{tikzpicture}
  \caption{A $4$-colored graph $\mathcal{G}$
and the 3-bubble $\mathcal{B}^{\hat{1}}_{ v \bar{v} }$ obtained by removing all edges of color $1$.}
  \label{G and bubble}
\end{figure}

Figure \ref{G and bubble} illustrates, for the case $D=3$, a $(D+1)-$colored graph $\mathcal{G}$ together with one of its 3-bubbles, $\mathcal{B}^{\hat{1}}_{ v \bar{v} }$.

\subsection{Colored graphs and the simplicial complexes of the dual pseudomanifolds}

Next, we examine how the simplicial complex dual to a $(D+1)$-colored graph $\mathcal{G}$ is defined through the bubbles of $\mathcal{G}$.
We begin by looking at what simplex is dual to a $(D+1-d)$-bubble of $\mathcal{G}$.
Consider the set $A$ of $D$-bubbles $\mathcal{B}^{\hat{i}}_{(\rho)}$ of $\mathcal{G}$:
\begin{align}
    A = \left\{ \mathcal{B}^{\hat{i}}_{(\rho)} \mid \forall \rho, i \in \{0, \dots, D\} \right\}.
\end{align}
Each $D$-bubble in this set corresponds to a 0-simplex.

Next, we show that a $(D+1-d)$-bubble is dual to a $(d-1)$-simplex.
Since a $(d-1)$-simplex is given as a set of $d$ 0-simplices,
it suffices to show that a $(D+1-d)$-bubble reproduces $d$ different $D$-bubbles.
Take a $(D+1-d)$-bubble $\mathcal{B}^{\hat{i}_1 \cdots \hat{i}_d}_{(\kappa)}$.
By adding to it the edges of colors
$i_1, i_2, \dots, i_{k-1}, i_{k+1}, \dots, i_d$
and the necessary new vertices 
so that it becomes a connected subgraph of $\mathcal{G}$,
we obtain a $D$-bubble $\mathcal{B}^{\hat{i}_k}_{(\rho)}$.
The $D$-bubble $\mathcal{B}^{\hat{i}_k}_{(\rho)}$ obtained from $\mathcal{B}^{\hat{i}_1 \cdots \hat{i}_d}_{(\kappa)}$
by this procedure is uniquely determined.
Since there are $d$ choices for the $(d-1)$ colors added to $\mathcal{B}^{\hat{i}_1 \cdots \hat{i}_d}_{(\kappa)}$,
we can associate $d$ different $D$-bubbles to any $(D+1-d)$-bubble $\mathcal{B}^{\hat{i}_1 \cdots \hat{i}_d}_{(\kappa)}$.
The resulting set of $D$-bubbles is written as
\begin{align}
    \sigma^{\mathcal{B}^{\hat{i}_1 \cdots \hat{i}_d}_{(\kappa)}}
    = \left\{
        \mathcal{B}^{\hat{i}_k}_{(\rho)}
        \mid
        \mathcal{B}^{\hat{i}_1 \cdots \hat{i}_d}_{(\kappa)}
        \subset
        \mathcal{B}^{\hat{i}_k}_{(\rho)},
        \ k \in \{1, \dots, d\}
      \right\}.
\end{align}
Since a $D$-bubble is dual to a 0-simplex,
the set $\sigma^{\mathcal{B}^{\hat{i}_1 \cdots \hat{i}_d}_{(\kappa)}}$
is a collection of $d$ 0-simplices and therefore forms a $(d-1)$-simplex.
This shows that a $(D+1-d)$-bubble is dual to a $(d-1)$-simplex.
Thus, given a $(D+1)$-colored graph,
one obtains all simplices from 0-simplices to $D$-simplices as those dual to its bubbles.

Next, we define the set $\Delta$
consisting of all simplices obtained as duals of the bubbles of $\mathcal{G}$:
\begin{align}
    \Delta \equiv \left\{
        \sigma^{\mathcal{B}^{\hat{i}_1 \cdots \hat{i}_d}_{(\kappa)}}
        \mid
        \mathcal{B}^{\hat{i}_1 \cdots \hat{i}_d}_{(\kappa)} \subset \mathcal{G}
    \right\}.
\end{align}
We will show that this is a simplicial complex.
The definition of a simplicial complex of a pseudomanifold is as follows:
\begin{dfn}
A simplicial complex of a pseudomanifold is a finite collection of simplices in $\mathbb{R}^D$
such that every face $\tau$ of any simplex $\sigma \in K$ also belongs to $K$.
That is, if $\sigma \in K$ and $\tau \le \sigma$, then $\tau \in K$.
\end{dfn}

A face of a $(d-1)$-simplex
$\sigma^{\mathcal{B}^{\hat{i}_1 \cdots \hat{i}_d}_{(\kappa)}} \in \Delta$
is a subset of it,
so it suffices to show that any subset $\tau$ of
$\sigma^{\mathcal{B}^{\hat{i}_1 \cdots \hat{i}_d}_{(\kappa)}}$
is also an element of $\Delta$.
Such a subset $\tau$ can be written, using a subset $S \subset \{1,2,\dots,d\}$, as:
\begin{align}
    \tau = \left\{
        \mathcal{B}^{\hat{i}_k}_{(\rho)}
        \mid
        \mathcal{B}^{\hat{i}_1 \cdots \hat{i}_d}_{(\kappa)}
        \subset
        \mathcal{B}^{\hat{i}_k}_{(\rho)},
        \ k \in \{1, \dots, d\} \backslash S
    \right\}.
\end{align}
This is the set of $D$-bubbles obtained from
$\mathcal{B}^{\hat{i}_1 \cdots \hat{i}_d}_{(\kappa)}$
by adding all edges of colors not equal to $i_k$ for
$k \in \{1, \dots, d\} \backslash S$.

Alternatively, $\tau$ can also be obtained by the following procedure.
First, add to $\mathcal{B}^{\hat{i}_1 \cdots \hat{i}_d}_{(\kappa)}$
all edges of colors $i_k$ for $k \in S$ using the same method.
Without loss of generality, by relabeling indices,
we may take $S = \{d-|S|+1, \dots, d\}$.
This yields a unique $(D+1 - d + |S|)$-bubble
\[
\mathcal{B}^{\hat{i}_1 \cdots \hat{i}_{d-|S|}}_{(\xi)}
\supset
\mathcal{B}^{\hat{i}_1 \cdots \hat{i}_d}_{(\kappa)}.
\]
Next, by adding to
$\mathcal{B}^{\hat{i}_1 \cdots \hat{i}_{d-|S|}}_{(\xi)}$
all edges whose colors are not $i_k$ for
$k \in \{1, \dots, d\} \backslash S=\{1,\dots,d-|S|\}$,
one obtains exactly one $D$-bubble for each added color.
The resulting set of such $D$-bubbles is precisely $\tau$.
Thus,
\begin{align}
    \tau =
    \sigma^{\mathcal{B}^{\hat{i}_1 \cdots \hat{i}_{d-|S|}}_{(\xi)}}
    = \left\{
        \mathcal{B}^{\hat{i}_k}_{(\rho)}
        \mid
        \mathcal{B}^{\hat{i}_1 \cdots \hat{i}_{d-|S|}}_{(\xi)}
        \subset
        \mathcal{B}^{\hat{i}_k}_{(\rho)},
        \ k \in \{1, \dots, d\} \backslash S
      \right\}.
\end{align}
By the definition of $\Delta$,
$\tau = \sigma^{\mathcal{B}^{\hat{i}_1 \cdots \hat{i}_{d-|S|}}_{(\xi)}}$ belongs to $\Delta$,
so $\Delta$ is a simplicial complex.

Finally, $\Delta$ satisfies the following three properties,
which are necessary and sufficient for being a pseudomanifold
\cite{Gurau:2011xp}:
\begin{enumerate}
    \item Every $(D-1)$-simplex is a face of exactly two $D$-simplices (\textit{non-branching}).
    \item Any two $D$-simplices sharing a $(D-1)$-simplex are so-called connected by a strong chain.
          By following such chains, any two $D$-simplices can be connected
          (\textit{strongly connected}).
    \item Every simplex is a face of some $D$-simplex (\textit{pure}).
\end{enumerate}

\subsection{Concrete example of analyzing a colored graph and its dual pseudomanifold}

\begin{figure}
  \centering
  \includegraphics[width=125mm]{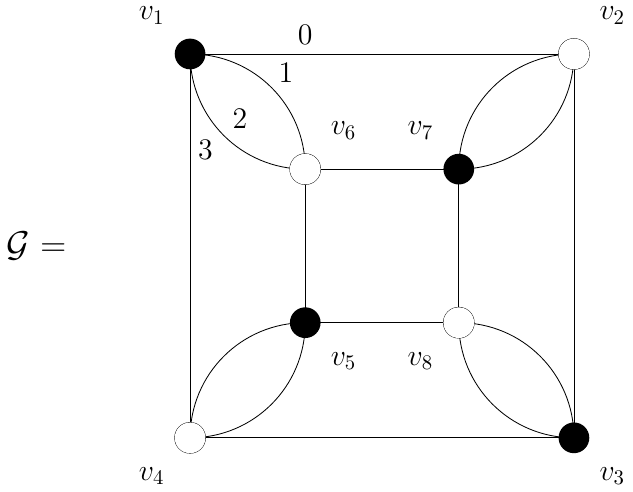}
  \caption{A 4-colored graph $\mathcal{G}$ consisting of eight vertices $v_1,\dots,v_8$ and edges colored $0,1,2,3$.}
  \label{3sphere}
\end{figure}

Given a $(D+1)$-colored graph $\mathcal{G}$, we can construct the pseudomanifold dual to $\mathcal{G}$ by determining the simplices corresponding to all bubbles of $\mathcal{G}$.  
As a concrete example of this procedure, in this subsection we analyze which pseudomanifold is dual to the 4-colored graph shown in Figure~\ref{3sphere}.

\begin{figure}
  \centering
  \includegraphics[width=125mm]{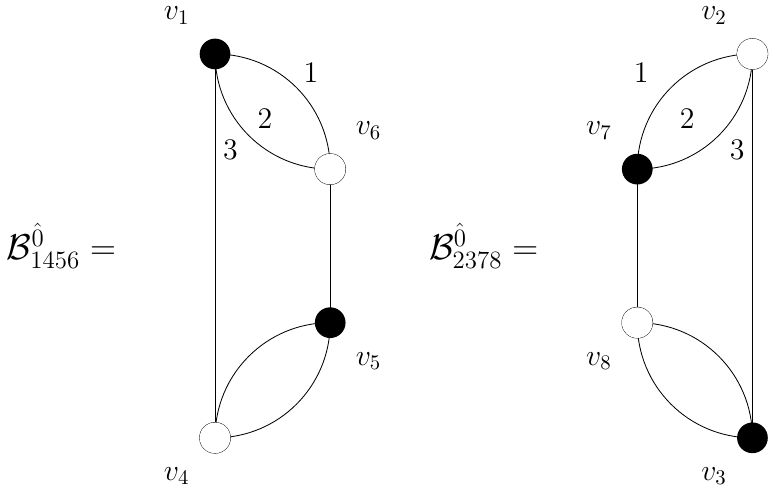}
  \caption{Two examples of $\mathcal{B}^{\hat{0}}_{v_1 v_2 \dots v_n}$ obtained from $\mathcal{G}$.  
  Removing all edges of a given color may cause the graph to decompose into multiple connected components.}
  \label{3bubble}
\end{figure}

We first examine the 3-bubbles $B^{\hat{i}_1}_{v_1 v_2 \dots v_n}$ of $\mathcal{G}$.  
A 3-bubble $B^{\hat{i}_1}_{v_1 v_2 \dots v_n}$ is defined as the connected subgraph obtained by removing from $\mathcal{G}$ all edges of color $i_1$ and taking the connected component containing the vertices $v_1,\dots,v_n$.  
Each such 3-bubble is dual to a 0-simplex, i.e., a vertex $A_k$ of the pseudomanifold.  
As explicit examples, Figure~\ref{3bubble} shows two bubbles $\mathcal{B}^{\hat{0}}_{v_1 v_2 \dots v_n}$ obtained from $\mathcal{G}$.
From $\mathcal{G}$ we obtain the following six 3-bubbles:
\begin{align}
	\mathcal{B}_{1456}^{\hat{0}} = A_1,\quad 
	\mathcal{B}_{2378}^{\hat{0}} = A_2,\quad
	\mathcal{B}_{12345678}^{\hat{1}} = A_3,\quad
	\mathcal{B}_{12345678}^{\hat{2}} = A_4,\quad
	\mathcal{B}_{1267}^{\hat{3}} = A_5,\quad
	\mathcal{B}_{3458}^{\hat{3}} = A_6 .
\end{align}
In what follows, the indices appearing as subscripts in the symbol $\mathcal{B}$ are always written in ascending numerical order, and simplices of the same dimension are assigned consecutive index numbers.

Next, we analyze the 2-bubbles $B^{\hat{i}_1 \hat{i}_2}_{v_1 v_2 \dots v_n}$ obtained from $\mathcal{G}$.  
These correspond to 1-simplices $\{A_k, A_l\}$ in the dual pseudomanifold.  
From $\mathcal{G}$ we obtain the following fourteen 2-bubbles:
\begin{align}
\begin{split}
	\mathcal{B}_{1456}^{\hat{0}\hat{1}} &= \{ A_1 , A_3 \}_1 (1, 3),\quad
	\mathcal{B}_{2378}^{\hat{0}\hat{1}} = \{ A_2 , A_3 \}_2 (2, 3),\quad
	\mathcal{B}_{1456}^{\hat{0}\hat{2}} = \{ A_1 , A_4 \}_3 (1, 4),\quad
	\mathcal{B}_{2378}^{\hat{0}\hat{2}} = \{ A_2 , A_4 \}_4 (2, 4),\\[6pt]
	\mathcal{B}_{16}^{\hat{0}\hat{3}} &= \{ A_1 , A_5 \}_5 (1, 5),\quad
	\mathcal{B}_{27}^{\hat{0}\hat{3}} = \{ A_2 , A_5 \}_6 (2, 5),\quad
	\mathcal{B}_{38}^{\hat{0}\hat{3}} = \{ A_2 , A_6 \}_7 (2, 6),\quad
	\mathcal{B}_{1234}^{\hat{1}\hat{2}} = \{ A_3 , A_4 \}_8 (3, 4),\\[6pt]
	\mathcal{B}_{5678}^{\hat{1}\hat{2}} &= \{ A_3 , A_4 \}_9 (3, 4),\quad
	\mathcal{B}_{1267}^{\hat{1}\hat{3}} = \{ A_3 , A_5 \}_{10} (3, 5),\quad
	\mathcal{B}_{3458}^{\hat{1}\hat{3}} = \{ A_3 , A_6 \}_{11} (3, 6),\\[6pt]
	\mathcal{B}_{1267}^{\hat{2}\hat{3}} &= \{ A_4 , A_5 \}_{12} (4, 5),\quad
	\mathcal{B}_{3458}^{\hat{2}\hat{3}} = \{ A_4 , A_6 \}_{13} (4, 6),\quad
	\mathcal{B}_{45}^{\hat{0}\hat{3}} = \{ A_1 , A_6 \}_{14} (1, 6).
\end{split}
\end{align}
In this way, in what follows, we attach to each $p$-simplex the information of the consecutive index numbers of the $(p-1)$-simplices that constitute it.
Here we explain how a 2-bubble corresponds to a 1-simplex, using  
\[
\mathcal{B}_{1456}^{\hat{0}\hat{1}} = \{A_1, A_3\}_1 (1,3)
\]
as a representative example.
First, consider adding an edge of color $0$ to the 2-bubble  
$\mathcal{B}_{1456}^{\hat{0}\hat{1}}$ so as to obtain a 3-bubble.  
The resulting 3-bubble must be one lacking color $1$, that is, a bubble of the form  
$\mathcal{B}^{\hat{1}}_{(\rho)}$.  
Among the 3-bubbles in $\mathcal{G}$ lacking color $1$, the only such bubble is
\[
\mathcal{B}^{\hat{1}}_{12345678} = A_3 .
\]
Thus, the simplex dual to $\mathcal{B}_{1456}^{\hat{0}\hat{1}}$ includes $A_3$.
Next, consider adding an edge of color $1$ to the same 2-bubble, thereby producing a 3-bubble lacking color $0$.  
The 3-bubbles of $\mathcal{G}$ lacking color $0$ are
\[
\mathcal{B}^{\hat{0}}_{1456} = A_1, 
\qquad 
\mathcal{B}^{\hat{0}}_{2378} = A_2 .
\]
However, $\mathcal{B}^{\hat{0}}_{2378}$ does not contain  
$\mathcal{B}_{1456}^{\hat{0}\hat{1}}$ as a subgraph and therefore cannot occur.  
Thus, the simplex dual to $\mathcal{B}_{1456}^{\hat{0}\hat{1}}$ includes $A_1$.
Consequently, the two 3-bubbles $\mathcal{B}_{1456}^{\hat{0}}$ and  
$\mathcal{B}_{12345678}^{\hat{1}}$ are dual to the two vertices $A_1$ and $A_3$, respectively,  
and the 2-bubble $\mathcal{B}_{1456}^{\hat{0}\hat{1}}$ is dual to the 1-simplex  
\[
\{A_1, A_3\}_1 (1, 3).
\]

From the graph $\mathcal{G}$ we obtain the following sixteen 1-bubbles, each of which
is dual to a 2-simplex:
\[
\begin{aligned}
\mathcal{B}_{14}^{\hat{0}\hat{1}\hat{2}} &= \{A_1, A_3, A_4\}_1 (1,3,8), 
&\mathcal{B}_{23}^{\hat{0}\hat{1}\hat{2}} &= \{A_2, A_3, A_4\}_2 (2,4,8),\\
\mathcal{B}_{56}^{\hat{0}\hat{1}\hat{2}} &= \{A_1, A_3, A_4\}_3 (1,3,9), 
&\mathcal{B}_{78}^{\hat{0}\hat{1}\hat{2}} &= \{A_2, A_3, A_4\}_4 (2,4,9),\\
\mathcal{B}_{16}^{\hat{0}\hat{1}\hat{3}} &= \{A_1, A_3, A_5\}_5 (1,5,10), 
&\mathcal{B}_{27}^{\hat{0}\hat{1}\hat{3}} &= \{A_2, A_3, A_5\}_6 (2,6,10),\\
\mathcal{B}_{38}^{\hat{0}\hat{1}\hat{3}} &= \{A_2, A_3, A_6\}_7 (2,7,11), 
&\mathcal{B}_{45}^{\hat{0}\hat{1}\hat{3}} &= \{A_1, A_3, A_6\}_8 (1,11,14),\\
\mathcal{B}_{16}^{\hat{0}\hat{2}\hat{3}} &= \{A_1, A_4, A_5\}_9 (3,5,12), 
&\mathcal{B}_{27}^{\hat{0}\hat{2}\hat{3}} &= \{A_2, A_4, A_5\}_{10} (4,6,12),\\
\mathcal{B}_{38}^{\hat{0}\hat{2}\hat{3}} &= \{A_2, A_4, A_6\}_{11} (4,7,13), 
&\mathcal{B}_{45}^{\hat{0}\hat{2}\hat{3}} &= \{A_1, A_4, A_6\}_{12} (3,13,14),\\
\mathcal{B}_{12}^{\hat{1}\hat{2}\hat{3}} &= \{A_3, A_4, A_5\}_{13} (8,10,12), 
&\mathcal{B}_{34}^{\hat{1}\hat{2}\hat{3}} &= \{A_3, A_4, A_6\}_{14} (8,11,13),\\
\mathcal{B}_{67}^{\hat{1}\hat{2}\hat{3}} &= \{A_3, A_4, A_5\}_{15} (9,10,12), 
&\mathcal{B}_{58}^{\hat{1}\hat{2}\hat{3}} &= \{A_3, A_4, A_6\}_{16} (9,11,13).
\end{aligned}
\]
From the 4-colored graph $\mathcal{G}$, we further obtain the following four
0-bubbles, each of which is dual to a 3-simplex:
\begin{align}
	\begin{split}
		\mathcal{B}_{1}^{\hat{0}\hat{1}\hat{2}\hat{3}}  &= \{ A_1, A_3, A_4, A_5 \}_1 (1, 5, 9, 13) , \mathcal{B}_{2}^{\hat{0}\hat{1}\hat{2}\hat{3}} = \{ A_2, A_3, A_4, A_5 \}_2 (2, 6, 10, 13), \\
		\mathcal{B}_{3}^{\hat{0}\hat{1}\hat{2}\hat{3}} &= \{ A_2, A_3, A_4, A_6 \}_3 (2, 7, 11, 14) , \ \mathcal{B}_{4}^{\hat{0}\hat{1}\hat{2}\hat{3}} = \{ A_1, A_3, A_4, A_6 \}_4 (1, 8, 12, 14), \\
		\mathcal{B}_{5}^{\hat{0}\hat{1}\hat{2}\hat{3}} &= \{ A_1, A_3, A_4, A_6 \}_5 (3, 8, 12, 16) , \ \mathcal{B}_{6}^{\hat{0}\hat{1}\hat{2}\hat{3}} = \{ A_1, A_3, A_4, A_5 \}_6 (3, 5, 9, 15), \\
		\mathcal{B}_{7}^{\hat{0}\hat{1}\hat{2}\hat{3}} &= \{ A_2, A_3, A_4, A_5 \}_7 (4, 6, 10, 15) , \ \mathcal{B}_{8}^{\hat{0}\hat{1}\hat{2}\hat{3}} = \{ A_2, A_3, A_4, A_6 \}_8 (4, 7, 11, 16).
	\end{split}
\end{align}

Summarizing the above, from the 4-colored graph $\mathcal{G}$ we have constructed
the following simplicial complex $\Delta$:
\begin{align}
	\begin{split}
		\Delta = \{ &A_1, A_2, A_3, A_4, A_5, A_6, 	 \\  
& \{ A_1 , A_3 \}_1 (1, 3) , \ 
\{ A_2 , A_3 \}_2 (2, 3), \
	 \{ A_1 , A_4 \}_3 (1, 4) , \ 
	  \{ A_2 , A_4 \}_4 (2, 4), \
	 \{ A_1 , A_5 \}_5 (1, 5),  \\
	&\{ A_2 , A_5 \}_6 (2, 5), \ 
	 \{ A_2 , A_6 \}_7 (2, 6), \
	 \{ A_3 , A_4 \}_8 (3, 4), 
	  \{ A_3 , A_4 \}_9 (3, 4), \
	 \{ A_3 , A_5 \}_{10} (3, 5), \\ 
	 &\{ A_3 , A_6 \}_{11} (3, 6), \ 
	 \{ A_4 , A_5 \}_{12} (4, 5), \
	  \{ A_4 , A_6 \}_{13} (4, 6), \ 
	  \{ A_1 , A_6 \}_{14} (1, 6), \\
	  & \{ A_1, A_3, A_4\}_1 (1, 3, 8),\ 
	   \{ A_2, A_3, A_4\}_2 (2, 4, 8),\
		 \{ A_1, A_3, A_4\}_3 (1, 3, 9), \\ 
		& \{ A_2, A_3, A_4\}_4 (2, 4, 9),\
		 \{ A_1, A_3, A_5 \}_5 (1, 5, 10), \ 
		  \{ A_2, A_3, A_5 \}_6 (2, 6, 10), \\
		& \{ A_2, A_3, A_6 \}_7 (2, 7, 11), \ \{ A_1, A_3, A_6 \}_8 (1, 11, 14), 
		\{ A_1, A_4, A_5 \}_9 (3, 5, 12), \\ &\{ A_2, A_4, A_5 \}_{10} (4, 6, 12), \
		 \{ A_2, A_4, A_6 \}_{11} (4, 7, 13), \ 
		 \{ A_1, A_4, A_6 \}_{12} (3, 13, 14), \\
 & \{ A_3, A_4, A_5 \}_{13} (8, 10, 12), \ 
  \{ A_3, A_4, A_6 \}_{14} (8, 11, 13), \
		 \{ A_3, A_4, A_5 \}_{15} (9, 10, 12), \\ 
 & \{ A_3, A_4, A_6 \}_{16} (9, 11, 13), \\
	&\{ A_1, A_3, A_4, A_5 \}_1 (1, 5, 9, 13) , 
	 \{ A_2, A_3, A_4, A_5 \}_2 (2, 6, 10, 13), \\
 &\{ A_2, A_3, A_4, A_6 \}_3 (2, 7, 11, 14) , \ 
  \{ A_1, A_3, A_4, A_6 \}_4 (1, 8, 12, 14), \\
& \{ A_1, A_3, A_4, A_6 \}_5 (3, 8, 12, 16) , \ 
  \{ A_1, A_3, A_4, A_5 \}_6 (3, 5, 9, 15), \\
& \{ A_2, A_3, A_4, A_5 \}_7 (4, 6, 10, 15) , \ 
 \{ A_2, A_3, A_4, A_6 \}_8 (4, 7, 11, 16)
		\}.
	\end{split}	\label{scs3}
\end{align}

\begin{figure}[h]
  \centering
  \includegraphics[width=100mm]{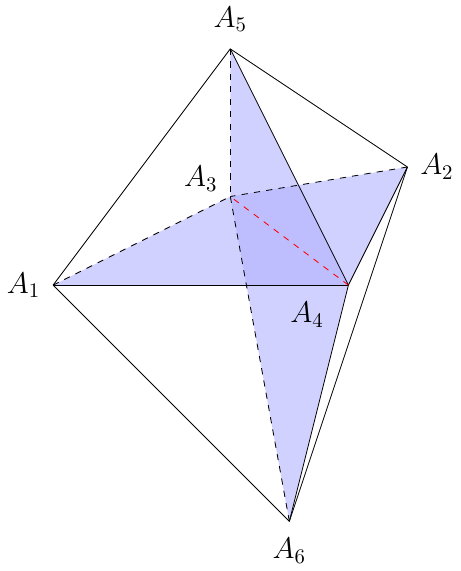}
  \caption{A manifold obtained by gluing two octahedra along their boundaries.  
  The colored faces and edges inside each octahedron are not glued together.}
  \label{8mentai}
\end{figure}

Next, we examine what kind of pseudomanifold the 3-dimensional simplicial complex $\Delta$ forms.
By focusing on the eight 3-simplices of $\Delta$ and using the information of the internal $(p-1)$-simplices attached to each $p$-simplex, we find that
\begin{align}
    \begin{split}
    &\{ A_1, A_3, A_4, A_5 \}_1 (1, 5, 9, 13),
    \{ A_2, A_3, A_4, A_5 \}_2 (2, 6, 10, 13), \\
    &\{ A_2, A_3, A_4, A_6 \}_3 (2, 7, 11, 14),
    \{ A_1, A_3, A_4, A_6 \}_4 (1, 8, 12, 14)
    \end{split} \label{scs4}
\end{align}
constitute a simplicial complex of an octahedron $A_1A_2A_3A_4A_5A_6$ that contains an interior (Fig.~\ref{8mentai}).  
On the other hand,
\begin{align}
    \begin{split}
    &\{ A_1, A_3, A_4, A_6 \}_5 (3, 8, 12, 16),
    \{ A_1, A_3, A_4, A_5 \}_6 (3, 5, 9, 15), \\
    &\{ A_2, A_3, A_4, A_5 \}_7 (4, 6, 10, 15),
    \{ A_2, A_3, A_4, A_6 \}_8 (4, 7, 11, 16)
    \end{split} \label{scs5}
\end{align}
also form a simplicial complex of an octahedron $A_1A_2A_3A_4A_5A_6$ with another interior.

The 1-simplices and 2-simplices on the boundary surface of these two octahedra are identical.  
However, the internal 1-simplex $\{ A_3, A_4 \}_8 (3, 4)$ of the former and  
the internal 1-simplex $\{ A_3, A_4 \}_9 (3, 4)$ of the latter are different.  
Moreover, the internal 2-simplices of the former,
\[
\{ A_1, A_3, A_4 \}_1 (1, 3, 8),\quad
\{ A_2, A_3, A_4 \}_2 (2, 4, 8),\quad
\{ A_3, A_4, A_5 \}_{13} (8, 10, 12),\quad
\{ A_3, A_4, A_6 \}_{14} (8, 11, 13),
\]
and the internal 2-simplices of the latter,
\[
\{ A_1, A_3, A_4 \}_3 (1, 3, 9),\quad
\{ A_2, A_3, A_4 \}_4 (2, 4, 9),\quad
\{ A_3, A_4, A_5 \}_{15} (9, 10, 12),\quad
\{ A_3, A_4, A_6 \}_{16} (9, 11, 13),
\]
are all distinct.
Each of the two octahedra with interiors is homeomorphic to a 3-ball.  
By identifying their boundary 2-spheres, we obtain a 3-sphere.  
Therefore, the 3-dimensional simplicial complex $\Delta$ is homeomorphic to the 3-sphere.

\section{Matrix representation of simplicial complexes}
\subsection{Correspondence between colored graphs and simplicial complex matrices}

In this paper, the first row of a matrix is referred to as the 0th row, and the following rows as the 1st row, 2nd row, and so on.  
The columns are referred to as the 1st column, 2nd column, and so on.

In the previous section, we reviewed that, given a $(D+1)$-colored graph $\mathcal{G}$, one can compute all bubbles of $\mathcal{G}$ and, by examining the relation between $(D-k)$-bubbles ($k = 1,2,\dots,D$) and $D$-bubbles, obtain the simplicial complex of the pseudomanifold dual to $\mathcal{G}$.

In this section, for the purpose of numerical computation, we introduce a representation of a $(D+1)$-colored graph by an integer matrix $M$.  
Using $M$, one can more easily compute all bubbles of a $(D+1)$-colored graph and analyze the relation between $(D-k)$-bubbles and the other $D$-bubbles.

Suppose that a $(D+1)$-colored graph $\mathcal{G}$ contains $|\mathcal{V}|$ vertices  
\[
v_1, v_2, \dots, v_{|\mathcal{V}|},
\]
where $v_{2n-1}$ is a positive vertex and $v_{2n}$ is a negative vertex for  
$n = 1,2,\dots, |\mathcal{V}|/2$.  
In this case, $\mathcal{G}$ can be represented by an integer matrix $M$ with $(D+1)$ rows and $|\mathcal{V}|$ columns.

The entry $M_{ij}$ in row $i$ and column $j$ is defined such that if an edge of color $i$ connects the two vertices $v_j$ and $v_k$, then  
\[
M_{ij} = k.
\]
By definition, $M_{ij} = k$ implies $M_{ik} = j$.  
Moreover, since edges connect only positive vertices to negative vertices, the parities of $M_{ij}$ and $j$ are different.  
When describing bubbles, if the vertex $v_j$ has no incident edge of color $i$, we set  
\[
M_{ij} = 0.
\]

Given the matrix $M$ corresponding to a $(D+1)$-colored graph $\mathcal{G}$, all $(D+1-d)$-bubbles of $\mathcal{G}$ can be found through the following procedure:

\begin{enumerate}
    \item In $M$, set all entries in the $i_1$-th,  $i_2$-th, $\dots$, $i_d$-th rows corresponding to the $d$ colors $i_1, i_2, \dots, i_d$ to zero.
    \item In the resulting matrix, identify subsets of  $k_1$-th, $k_2$-th, $\dots$, $k_{2l}$-th columns whose entries consist only of $0, k_1, k_2, \dots, k_{2l}$, and define submatrices from each such set of columns.  
    The resulting matrix is expressed as the sum of these submatrices, representing the direct sum decomposition into connected graphs.
    \item Each submatrix corresponding to a connected subgraph represents a $(D+1-d)$-bubble.
\end{enumerate}

The numerical algorithm for finding sets of columns that represent connected graphs is as follows.  
Let us consider the case where a matrix $M$ corresponding to a bubble of a general colored graph $\mathcal{G}$ is given.

\begin{enumerate}
    \item Among the columns of $M$, find one that contains at least one nonzero entry.  
        Suppose this is the $m$-th column.  
        Create a list $A=[m]$, which represents the list of columns.
    \item In the $m$-th column of $M$, suppose there are $k$ nonzero entries.  
        Let these entries be $m_1, m_2, \dots, m_k$, and construct a list  
        $B=[m_1, m_2, \dots, m_k]$, which represents the list of entries not yet included in $A$.
    \item  
    \begin{enumerate}
        \item Let the first element of list $B$ be $m_i$.  
            In the $m_i$-th column of $M$, suppose there are $l$ nonzero entries.  
            Let these be $n_1, n_2, \dots, n_l$.
            \begin{enumerate}
                \item If $n_j$ is already contained in list $A$, do nothing.  
                       If it is not contained in $A$, append $n_j$ to the end of list $B$.
                \item Perform the operation in 3(a)(i) for each $j = 1, 2, \dots, l$.
            \end{enumerate}
        \item Remove all elements in list $B$ that have the same value as $m_i$,  
              and append $m_i$ to the end of list $A$.
    \end{enumerate}
    \item Repeat step 3 until list $B$ becomes empty.  
        The resulting list $A = [a_1, a_2, \dots, a_n]$ represents the set of columns corresponding to a connected graph.
    \item Redefine $M$ as the matrix obtained by setting all entries in columns $a_1, a_2, \dots, a_n$ to zero.
    \item If the redefined matrix $M$ still contains nonzero entries, repeat steps 1 through 5.  
        If it contains no nonzero entries, terminate the procedure.
\end{enumerate}
Using this procedure, one can find all sets of columns in $M$ that represent connected graphs,  
and the matrix $M$ can be expressed as a sum of matrices, each representing a connected graph.

As an example, consider the following matrix $M$:
\begin{align}
    M =
    \begin{pmatrix}
    2 & 1 & 4 & 3 & 8 & 7 & 6 & 5 \\
    6 & 7 & 8 & 5 & 4 & 1 & 2 & 3 \\
    6 & 7 & 8 & 5 & 4 & 1 & 2 & 3 \\
    4 & 3 & 2 & 1 & 6 & 5 & 8 & 7 \\
    \end{pmatrix}.
\end{align}
This matrix represents the graph $\mathcal{G}$ shown in Fig.~\ref{3sphere}.
The graph $\mathcal{G}$ is connected, and indeed, according to the above procedure, the only tuple of columns representing a connected graph is the trivial one, $M$ itself.

Next, consider the matrix $M^{\hat{0}}$ obtained by setting all entries in the 0-th row of $M$ to zero:
\begin{align}
    M^{\hat{0}} =
    \begin{pmatrix}
    0 & 0 & 0 & 0 & 0 & 0 & 0 & 0 \\
    6 & 7 & 8 & 5 & 4 & 1 & 2 & 3 \\
    6 & 7 & 8 & 5 & 4 & 1 & 2 & 3 \\
    4 & 3 & 2 & 1 & 6 & 5 & 8 & 7 \\
    \end{pmatrix}.
\end{align}
This matrix represents the graph obtained from $\mathcal{G}$ by removing all edges of color 0.
Focusing on $1$-st, $4$-th, $5$-th and $6$-th  columns of $M^{\hat{0}}$, we see that their entries consist only of $1,4,5,6$.
Similarly, $2$-nd, $3$-rd, $7$-th and $8$-th  columns contain only $2,3,7,8$.
Therefore, $M^{\hat{0}}$ represents the direct sum of two connected graphs on vertices $\{v_1, v_4, v_5, v_6\}$ and $\{v_2, v_3, v_7, v_8\}$, respectively.
Hence, it may be rewritten as the sum of the matrices corresponding to those two graphs.
These matrices correspond to the 3-bubbles $\mathcal{B}^{\hat{0}}_{1456}$ and $\mathcal{B}^{\hat{0}}_{2378}$:
\begin{align}
    M^{\hat{0}} &=
    \begin{pmatrix}
    0 & 0 & 0 & 0 & 0 & 0 & 0 & 0 \\
    6 & 7 & 8 & 5 & 4 & 1 & 2 & 3 \\
    6 & 7 & 8 & 5 & 4 & 1 & 2 & 3 \\
    4 & 3 & 2 & 1 & 6 & 5 & 8 & 7 \\
    \end{pmatrix} \nonumber \\
    &=
    \begin{pmatrix}
    0 & 0 & 0 & 0 & 0 & 0 & 0 & 0 \\
    6 & 0 & 0 & 5 & 4 & 1 & 0 & 0 \\
    6 & 0 & 0 & 5 & 4 & 1 & 0 & 0 \\
    4 & 0 & 0 & 1 & 6 & 5 & 0 & 0 \\
    \end{pmatrix}
    +
    \begin{pmatrix}
    0 & 0 & 0 & 0 & 0 & 0 & 0 & 0 \\
    0 & 7 & 8 & 0 & 0 & 0 & 2 & 3 \\
    0 & 7 & 8 & 0 & 0 & 0 & 2 & 3 \\
    0 & 3 & 2 & 0 & 0 & 0 & 8 & 7 \\
    \end{pmatrix} \nonumber \\
    &\equiv \mathcal{B}^{\hat{0}}_{1456} + \mathcal{B}^{\hat{0}}_{2378} \ .
\end{align}

\begin{figure}[h]
    \centering
    \includegraphics[width=100mm]{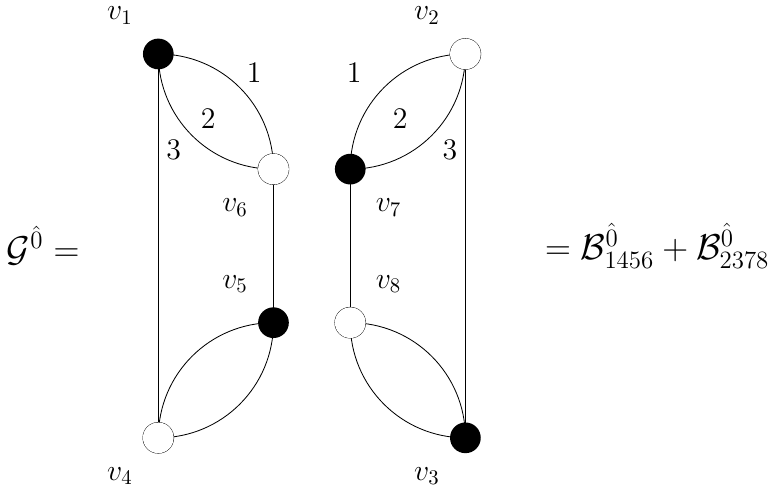}
    \caption{Removing all color-0 edges from $\mathcal{G}$ yields the direct sum of two connected subgraphs, which correspond to the 3-bubbles $\mathcal{B}^{\hat{0}}_{1456}$ and $\mathcal{B}^{\hat{0}}_{2378}$.}
    \label{0nasi}
\end{figure}

Indeed, in the actual graph, removing all edges of color 0 results in the direct sum of two connected subgraphs, each corresponding to the 3-bubbles $\mathcal{B}^{\hat{0}}_{1456}$ and $\mathcal{B}^{\hat{0}}_{2378}$ (Fig.~\ref{0nasi}).

As another example, consider the matrix obtained by removing all color-1 edges:
\begin{align}
    M^{\hat{1}} =
        \begin{pmatrix}
        2 & 1 & 4 & 3 & 8 & 7 & 6 & 5 \\
        0 & 0 & 0 & 0 & 0 & 0 & 0 & 0 \\
        6 & 7 & 8 & 5 & 4 & 1 & 2 & 3 \\
        4 & 3 & 2 & 1 & 6 & 5 & 8 & 7 \\
        \end{pmatrix}.
\end{align}
In this case, the only tuple of columns representing a connected graph is again the trivial one $M^{\hat{1}}$ itself.
Therefore, the subgraph obtained from $\mathcal{G}$ by removing all edges of color 1 is a single 3-bubble.
That is,
\[
    M^{\hat{1}} \equiv \mathcal{B}^{\hat{1}}_{12345678}.
\]

Moreover, by using this matrix $M$, it is possible to examine the relation between $D$-bubbles and $(D-k)$-bubbles.
Suppose that we are given the matrix 
$M^{\hat{i}_1,\hat{i}_2,\dots,\hat{i}_{k+1}}_{(\rho)}$ 
representing a $(D-k)$-bubble 
$\mathcal{B}^{\hat{i}_1,\hat{i}_2,\dots,\hat{i}_{k+1}}_{(\rho)}$, 
as well as the matrices 
$M^{\hat{i}_1}_{(\zeta_1)},M^{\hat{i}_2}_{(\zeta_2)}, \dots, M^{\hat{i}_{l}}_{(\zeta_{l})}$ 
corresponding to all $l$ distinct $D$-bubbles 
$\mathcal{B}^{\hat{i}_1}_{(\zeta_1)},\mathcal{B}^{\hat{i}_2}_{(\zeta_2)}, \dots, \mathcal{B}^{\hat{i}_{l}}_{(\zeta_{l})}$.
In this situation, among the matrices 
$M^{\hat{i}_1}_{(\zeta_1)},M^{\hat{i}_2}_{(\zeta_2)}, \dots, M^{\hat{i}_{l}}_{(\zeta_{l})}$, 
there always exist exactly $k+1$ matrices that contain all the non-zero entries of 
$M^{\hat{i}_1,\hat{i}_2,\dots,\hat{i}_{k+1}}_{(\rho)}$.

As a concrete example, consider the case in which the graph $\mathcal{G}$ is given by Figure~\ref{3sphere}.
We study the relation between the 1-bubble 
$\mathcal{B}^{\hat{0}\hat{2}\hat{3}}_{v_1 v_6}$ 
of $\mathcal{G}$ and its 3-bubbles by using the matrix $M$.
First, the matrix representing 
$\mathcal{B}^{\hat{0}\hat{2}\hat{3}}_{v_1 v_6}$ 
is
\begin{align}
    M^{\hat{0}\hat{2}\hat{3}}_{v_1 v_6} =
    \begin{pmatrix}
    0 & 0 & 0 & 0 & 0 & 0 & 0 & 0 \\
    6 & 0 & 0 & 0 & 0 & 1 & 0 & 0 \\
    0 & 0 & 0 & 0 & 0 & 0 & 0 & 0 \\
    0 & 0 & 0 & 0 & 0 & 0 & 0 & 0 \\
    \end{pmatrix}.
\end{align}
On the other hand, the matrices representing the six 3-bubbles of $\mathcal{G}$,
$\mathcal{B}_{1456}^{\hat{0}}$, $\mathcal{B}_{2378}^{\hat{0}}$, 
$\mathcal{B}_{12345678}^{\hat{1}}$,
$\mathcal{B}_{12345678}^{\hat{2}}$, 
$\mathcal{B}_{1267}^{\hat{3}}$ and
$\mathcal{B}_{3458}^{\hat{3}}$
are given respectively by
\begin{align}
    M_{1456}^{\hat{0}} &=
    \begin{pmatrix}
    0 & 0 & 0 & 0 & 0 & 0 & 0 & 0 \\
    6 & 0 & 0 & 5 & 4 & 1 & 0 & 0 \\
    6 & 0 & 0 & 5 & 4 & 1 & 0 & 0 \\
    4 & 0 & 0 & 1 & 6 & 5 & 0 & 0 \\
    \end{pmatrix}, \nonumber \\
    M_{2378}^{\hat{0}} &=
    \begin{pmatrix}
    0 & 0 & 0 & 0 & 0 & 0 & 0 & 0 \\
    0 & 7 & 8 & 0 & 0 & 0 & 2 & 3 \\
    0 & 7 & 8 & 0 & 0 & 0 & 2 & 3 \\
    0 & 3 & 2 & 0 & 0 & 0 & 8 & 7 \\
    \end{pmatrix}, \nonumber \\
    M_{12345678}^{\hat{1}} &=
    \begin{pmatrix}
    2 & 1 & 4 & 3 & 8 & 7 & 6 & 5 \\
    0 & 0 & 0 & 0 & 0 & 0 & 0 & 0 \\
    6 & 7 & 8 & 5 & 4 & 1 & 2 & 3 \\
    4 & 3 & 2 & 1 & 6 & 5 & 8 & 7 \\
    \end{pmatrix}, \nonumber \\
    M_{12345678}^{\hat{2}} &=
    \begin{pmatrix}
    2 & 1 & 4 & 3 & 8 & 7 & 6 & 5 \\
    6 & 7 & 8 & 5 & 4 & 1 & 2 & 3 \\
    0 & 0 & 0 & 0 & 0 & 0 & 0 & 3 \\
    4 & 3 & 2 & 1 & 6 & 5 & 8 & 7 \\
    \end{pmatrix}, \nonumber \\
    M_{1267}^{\hat{3}} &=
    \begin{pmatrix}
    2 & 1 & 0 & 0 & 0 & 7 & 6 & 0 \\
    6 & 7 & 0 & 0 & 0 & 1 & 2 & 0 \\
    6 & 7 & 0 & 0 & 0 & 1 & 2 & 0 \\
    0 & 0 & 0 & 0 & 0 & 0 & 0 & 0 \\
    \end{pmatrix}, \nonumber \\
    M_{3458}^{\hat{3}} &=
    \begin{pmatrix}
    0 & 0 & 4 & 3 & 8 & 0 & 0 & 5 \\
    0 & 0 & 8 & 5 & 4 & 0 & 0 & 3 \\
    0 & 0 & 8 & 5 & 4 & 0 & 0 & 3 \\
    0 & 0 & 0 & 0 & 0 & 0 & 0 & 7 \\
    \end{pmatrix}. \nonumber
\end{align}
Among these matrices representing the 3-bubbles, those that contain all the non-zero entries of 
$M^{\hat{0}\hat{2}\hat{3}}_{v_1 v_6}$ 
are  
$M_{1456}^{\hat{0}}$, $M_{12345678}^{\hat{2}}$ and $M_{1267}^{\hat{3}}$.
From this observation, we may conclude that the simplex dual to  
$\mathcal{B}^{\hat{0}\hat{2}\hat{3}}_{v_1 v_6}$ 
is composed of the three simplices dual to 
$\mathcal{B}_{1456}^{\hat{0}}$, 
 $\mathcal{B}_{12345678}^{\hat{2}}$ and
 $\mathcal{B}_{1267}^{\hat{3}}$.
In this way, the procedure for determining the bubbles of the graph $\mathcal{G}$  
and the relation between $D$-bubbles and $(D-k)$-bubbles  
can be systematically carried out by using the matrix $M$ associated with $\mathcal{G}$.

\subsection{Direct correspondence between simplicial complexes and simplicial-complex matrices}

Combining the correspondence between simplicial complexes and colored graphs, and the correspondence between colored graphs and simplicial-complex matrices, a simplicial-complex matrix is defined as a matrix satisfying the following two conditions:
(1) A $(D+1)\times 2N$ ($D$ and $N$ are natural numbers) matrix $M$ with natural number entries, and when $M_{ij}=k$, it satisfies $M_{ik}=j$.
(2) $M_{ij}+j \equiv 1 \pmod{2}$.

The code in Appendix B.1 randomly generates simplicial-complex matrices satisfying the above conditions when the dimension $D$ of the pseudomanifold is fixed.  
The code in Appendix B.3 performs the decomposition of a simplicial-complex matrix into its simplex matrices.  
The code in Appendix B.4 constructs the simplicial complex from a simplicial-complex matrix and its simplex matrices.  
The code in Appendix B.5, B.6, and B.7 computes, respectively, the volume of a simplex, the circumcenter of a simplex, and the volume of the dual simplex.

\section{Mutation and crossover of simplicial complexes}

In the previous section, we established the correspondence among simplicial complexes of pseudomanifolds, colored graphs, and simplicial-complex matrices.  
In this section, we transplant the concepts of mutation and crossover in genetic algorithms to colored graphs, and using the correspondence, define mutation and crossover of simplicial-complex matrices.  
By the correspondence, this defines mutation and crossover of the simplicial complexes themselves.

\subsection{Mutation and crossover of colored graphs}

\begin{figure}
    \centering
    \begin{tikzpicture}[scale=0.8, transform shape]
        \draw (0,0) to (4,0);
        \draw (0,4) to (4,4);

        \draw [color=red] (4,0) to (4,4);
        \draw [color=red] (0,0) to (0,4);
        
        \draw [color=blue, out=45, in=135] (0,0) to (4,0);
        \draw [color=blue, out=-45, in=-135] (0,4) to (4,4);

        \draw (0,0) circle [radius=0.2];
        \fill [white] (0,0) circle [radius=0.2];
        \draw (-0.5,-0.5) node [font=\Large] {$4$};

        \fill [black] (4,0) circle [radius=0.2];
        \draw (4.5,-0.5) node [font=\Large] {$3$};

        \fill [black] (0,4) circle [radius=0.2];
        \draw (-0.5,4.5) node [font=\Large] {$1$};

        \draw (4,4) circle [radius=0.2];
        \fill [white] (4,4) circle [radius=0.2];
        \draw (4.5,4.5) node [font=\Large] {$2$};

        \draw (8,0) to (12,0);
        \draw [color=red] (8,0) to (8,4);
        \draw [color=blue, out=45, in=-45] (8,0) to (8,4);
        \draw (8,4) to (12,4);
        \draw [color=red] (12,0) to (12,4);
        \draw [color=blue, out=135, in=-135] (12,0) to (12,4);

        \draw (8,0) circle [radius=0.2];
        \fill [white] (8,0) circle [radius=0.2];
        \draw (7.5,-0.5) node [font=\Large] {$4$};

        \draw (12,0) circle [radius=0.2];
        \fill [black] (12,0) circle [radius=0.2];
        \draw (12.5,-0.5) node [font=\Large] {$3$};

        \fill [black] (8,4) circle [radius=0.2];
        \draw (7.5,4.5) node [font=\Large] {$1$};

        \draw (12,4) circle [radius=0.2];
        \fill [white] (12,4) circle [radius=0.2];
        \draw (12.5,4.5) node [font=\Large] {$2$};

        \draw [<->, >=stealth, thick] (4.5,2) to (7.5,2);
        \draw (6,2.5) node [font=\Large] {Mutation};
    \end{tikzpicture}
    \caption{An example of mutation of a colored graph.}
    \label{graph_mutation}
\end{figure}
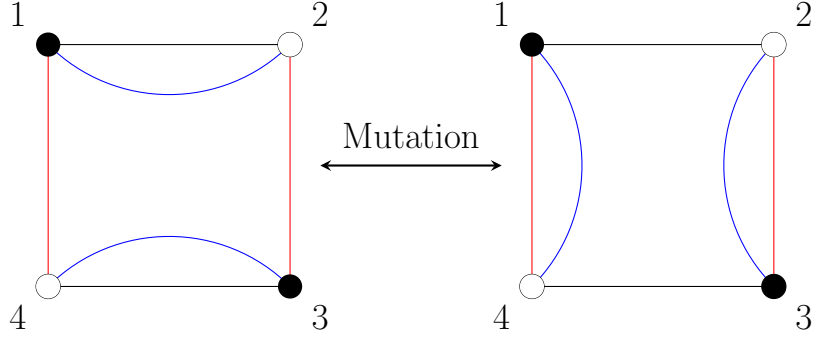

Mutation of a colored graph is a recombination of edges of the same color. More precisely,
\begin{dfn}[]
Given a positive vertex 1 and a negative vertex $1'$ connected by an edge of some color, and another positive vertex 2 and negative vertex $2'$ connected by another edge of the same color, we remove these two edges and add two new edges of the same color connecting vertex 1 to vertex $2'$ and vertex $1'$ to vertex 2. This operation is called a mutation of the colored graph.
\end{dfn}
A concrete example of mutation of a colored graph is given in Figure \ref{graph_mutation}.

\begin{figure}[h]
    \centering
    \includegraphics[width=100mm]{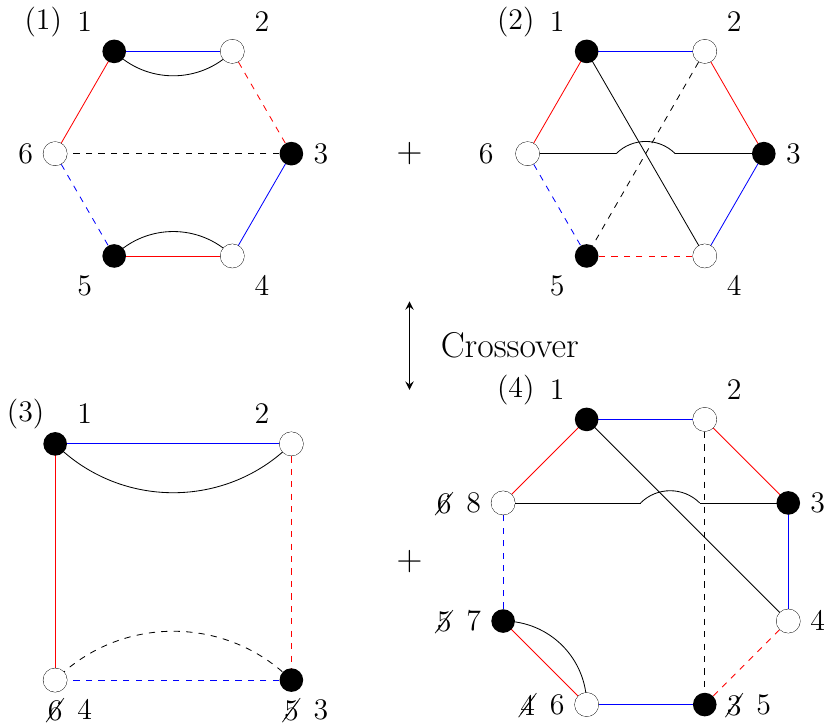}
    \caption{An example of crossover of colored graphs.}
    \label{graph_crossover}
\end{figure}

Crossover of colored graphs is performed by splitting two colored graphs and recombining the pieces. More precisely,
\begin{dfn}[]
In a $(D+1)$-colored graph, remove $(D+1)$ edges of mutually distinct colors, thereby splitting the graph into two connected components. At this moment, in one of the connected components, count the cumulative total number of  the positive and negative vertices that were incident to the removed edges. Then perform the same type of split on another $(D+1)$-colored graph so that the cumulative total number of positive and negative vertices in the corresponding component matches. Recombine the connected components, and reconnect each positive vertex and negative vertex using edges of the original colors. The two resulting graphs are called the crossover of the colored graphs.
\end{dfn}
A concrete example of crossover of colored graphs is given in Figure \ref{graph_crossover}.

\subsection{Mutation and crossover of simplicial complex matrices}

In this subsection, using the correspondence between colored graphs and simplicial complex matrices established in the previous section, we give the matrix representation of the mutation and crossover operations of colored graphs defined earlier.

We first define mutation for a single simplicial complex matrix.  
Choose a $i$-th row. This corresponds to selecting a color.  
Next, choose a $j$-th column. This corresponds to selecting a vertex $j$.  
Then $M_{ij}$ of opposite parity to $j$ is determined.  
This means that vertex $v_j$ is connected to vertex $v_{M_{ij}}$ by an edge of color $i$.  
In this case, the inverse relation $M_{i M_{ij}}=j$ also holds.

Choose another column $j'$ of the same parity as $j$.  
This corresponds to choosing another vertex $v_{j'}$.  
Then $M_{ij'}$ of opposite parity to $j'$ is determined,  
which means that vertex $v_{j'}$ is connected to vertex $v_{M_{ij'}}$ by an edge of color $i$.  
Here again the inverse relation $M_{i M_{ij'}}=j'$ holds.

\begin{figure}[h]
    \centering
    \begin{tikzpicture}[scale=0.8, transform shape]
        \matrix (m) [matrix of math nodes, nodes={minimum width=10mm, minimum height=7mm}, left delimiter={(}, right delimiter={)}]
        {
            2 & 1 & 4 & 3 \\
            4 & 3 & 2 & 1 \\
            2 & 1 & 4 & 3 \\
        };

        \foreach \col in {2,4} {
            \node (num\col) at ($(m-1-\col.north) + (0,4mm)$) {\col};
            \node[draw, circle, minimum size=0.5mm] at ($(num\col.north) + (0,2mm)$) {};
        }
        \foreach \col in {1,3} {
            \node (num\col) at ($(m-1-\col.north) + (0,4mm)$) {\col};
            \node[fill, circle, minimum size=0.5mm] at ($(num\col.north) + (0,2mm)$) {};
        }

        \draw [thick] (m-3-1.south west) rectangle (m-3-1.north east);
        \draw [thick] (m-3-3.south west) rectangle (m-3-3.north east);
        \draw [thick] (m-3-2) circle [radius=4mm];
        \draw [thick] (m-3-4) circle [radius=4mm];

        \draw [<->, thick, out=-45, in=-135] ($(m-3-1.south) + (0pt,-4pt)$) to ($(m-3-3.south) + (0pt,-4pt)$);
        \draw [<->, thick, out=-45, in=-135] ($(m-3-2.south) + (0pt,-4pt)$) to ($(m-3-4.south) + (0pt,-4pt)$);

        \matrix (m2) [matrix of math nodes, nodes={minimum width=10mm, minimum height=7mm}, left delimiter={(}, right delimiter={)}, right = 2.5cm of m]
            {
                2 & 1 & 4 & 3 \\
                4 & 3 & 2 & 1 \\
                4 & 3 & 2 & 1 \\
            };

        \foreach \col in {2,4} {
            \node (num\col) at ($(m2-1-\col.north) + (0,4mm)$) {\col};
            \node[draw, circle, minimum size=0.5mm] at ($(num\col.north) + (0,2mm)$) {};
        }
        \foreach \col in {1,3} {
            \node (num\col) at ($(m2-1-\col.north) + (0,4mm)$) {\col};
            \node[fill, circle, minimum size=0.5mm] at ($(num\col.north) + (0,2mm)$) {};
        }

        \draw [thick] (m2-3-1.south west) rectangle (m2-3-1.north east);
        \draw [thick] (m2-3-3.south west) rectangle (m2-3-3.north east);
        \draw [thick] (m2-3-2) circle [radius=4mm];
        \draw [thick] (m2-3-4) circle [radius=4mm];

        \draw[->, thick]
        ($(m.east) + (10pt,0)$) -- node[above]{Mutation} ($(m2.west) + (-10pt,0)$);
    \end{tikzpicture}
    \caption{Mutation of simplicial complex matrix corresponding to Figure \ref{graph_mutation}. First, second and third rows of matrices represent the black, red and blue edges of the color graph, respectively.}
    \label{matrix_mutation}
\end{figure}
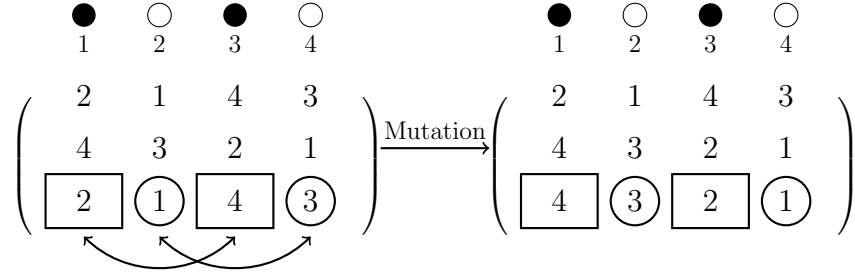

We now exchange the values $M_{ij}$ and $M_{ij'}$.  
In the resulting matrix, the $(i,j)$-entry becomes $M_{ij'}$, meaning that vertex $v_{j}$ is now connected to vertex $v_{M_{ij'}}$, and the $(i,j')$-entry becomes $M_{ij}$, meaning that vertex $v_{j'}$ is now connected to vertex $v_{M_{ij}}$.  
Therefore, since the inverse relations must also hold: $M_{i M_{ij'}}=j$ and $M_{i M_{ij}}=j'$, we must exchange the values of  
$M_{i M_{ij}}=j$ and $M_{i M_{ij'}}=j'$ as well.  
Because $j$ and $j'$, as well as $M_{ij}$ and $M_{ij'}$, have the same parity, this operation corresponds to a consistent recombination of positive and negative vertices of the edges in the colored graph.  
From these observations, the mutation of the matrix correctly represents the mutation of the colored graph.  
A concrete example of mutation of a simplicial complex matrix  corresponding to the example in Figure \ref{graph_mutation} is shown in Figure \ref{matrix_mutation}.

\vspace{1em}

Next, we define crossover between two simplicial complex matrices  
$M_{ij}$ and $N_{ik}$ ($i=0,\ldots,D$, $j=1,\ldots,2M$, $k=1,\ldots,2N$).  
We begin by defining a decomposition of the matrices.  
Consider the set of columns of $M$ as  
\[
\{ M_{i1}, M_{i2}, \ldots, M_{i\,2M} \} = \{ M_{iA} \}, \qquad (|A| = 2M).
\]
We divide this set into two disjoint subsets
\[
\{ M_{iA} \} = \{ M_{i\tilde{A}} \} + \{ M_{i\tilde{\tilde{A}}} \},
\qquad (|\tilde{A}| + |\tilde{\tilde{A}}| = 2M).
\]
The division is chosen so that, except for those components satisfying  
$M_{i\tilde{A}}=\tilde{A}'$,  
the subset $\{ M_{i\tilde{A}} \}$ contains exactly the $D+1$ entries
\[
M_{0\tilde{A}_0}, M_{1\tilde{A}_1}, \ldots, M_{D\tilde{A}_D},
\]
all with different row indices.  
This corresponds to the connected component of the colored graph obtained by removing the $D+1$ edges (one for each color) that connect $\{ M_{i\tilde{A}} \}$ to $\{ M_{i\tilde{\tilde{A}}} \}$.  
Since the original colored graph is connected, the same holds for the decomposition of $\{ M_{i\tilde{\tilde{A}}} \}$.
We perform an analogous decomposition for $N_{ik}$:
\[
\{ N_{i1}, N_{i2}, \ldots, N_{i\,2N} \} = \{ N_{iB} \}, \qquad (|B| = 2N),
\]
\[
\{ N_{iB} \} = \{ N_{i\tilde{B}} \} + \{ N_{i\tilde{\tilde{B}}} \},
\qquad (|\tilde{B}| + |\tilde{\tilde{B}}| = 2N),
\]
but we choose the decomposition so that the number of even and odd numbers among  
\[
M_{0\tilde{A}_0}, M_{1\tilde{A}_1}, \ldots, M_{D\tilde{A}_D}
\]
matches the number of odd and even numbers among  
\[
N_{0\tilde{B}_0}, N_{1\tilde{B}_1}, \ldots, N_{D\tilde{B}_D},
\] respectively.
This corresponds to ensuring that the number of positive and negative vertices connected to the removed edges in one colored graph matches those in the other.

\begin{figure}[h]
    \centering
    \includegraphics[width=100mm]{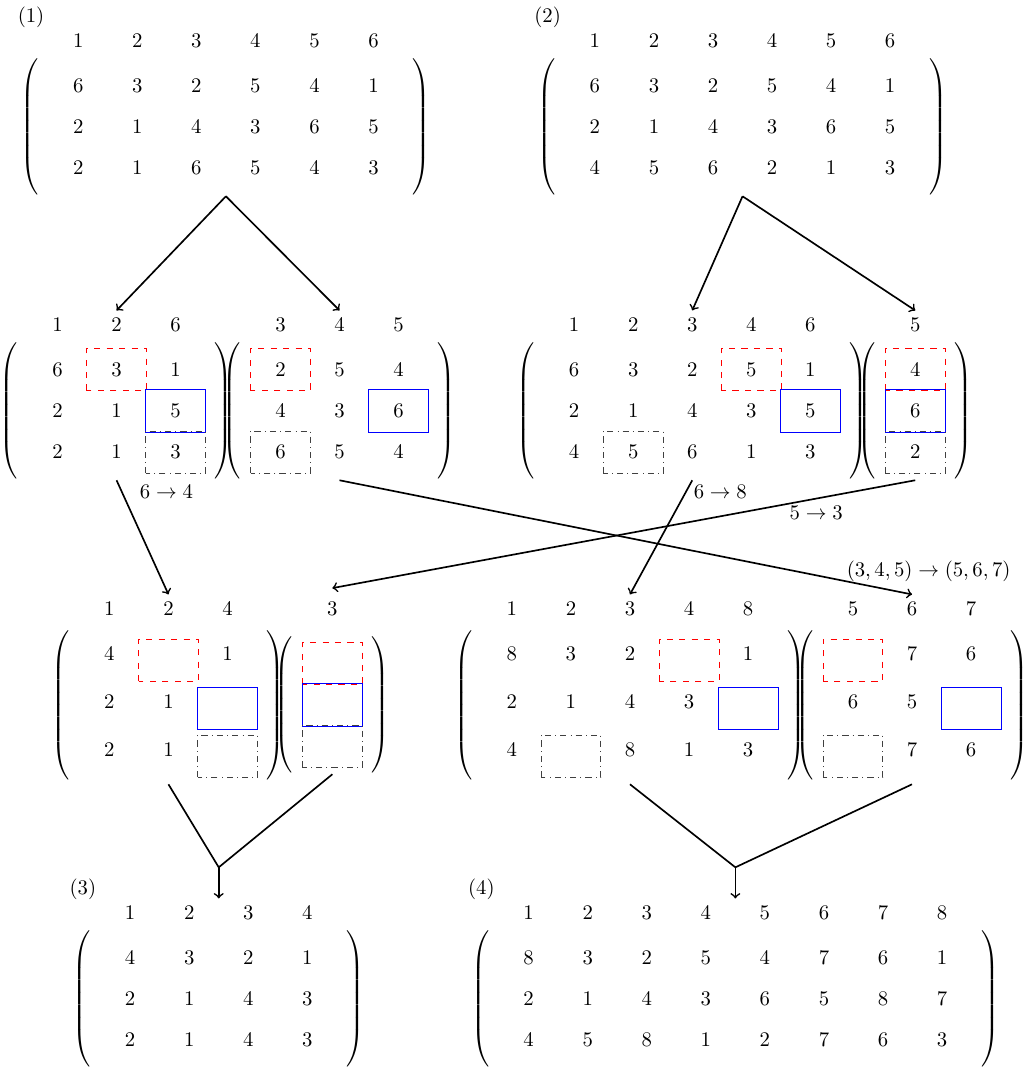}
    \caption{Mutation of a simplicial complex matrix corresponding to Figure \ref{graph_crossover}.}
    \label{matrix_crossover}
\end{figure}

We now define the recombination of the matrices.  
We relabel the sets $\{ M_{i\tilde{A}} \}$ and $\{ N_{i\tilde{B}} \}$ so that their labels $\tilde{A}$ and $\tilde{B}$ are exchanged while preserving parity, thereby combining them into a single matrix  
$\{ M^{new}_{i\tilde{A}_{new}} \}$.  
At the same time, the values of $M_{i\tilde{A}}$ and $N_{i\tilde{B}}$ are exchanged according to their original assignments  
$M_{i\tilde{A}}=\tilde{A}'$, $N_{i\tilde{B}}=\tilde{B}'$.
To recombine the matirices, we redefine  the $D+1$ entries:
\[
M^{new}_{i\tilde{A}^{new}_i} := \tilde{B}^{new}_i,
\qquad
N^{new}_{i\tilde{B}^{new}_i} := \tilde{A}^{new}_i.
\]
Here, $N^{new}_{i\tilde{B}^{new}_i}$ corresponds to entries originally belonging to $N_{i\tilde{B}_i}$ but relabeled within $\{ M^{new}_{i\tilde{A}_{new}} \}$.  
This operation corresponds to reconnecting the vertices from which edges were removed.  
The remaining entries $\{ N^{new}_{i\tilde{\tilde{B}}_{new}} \}$ are defined analogously.  
In this way, we obtain the matrices after crossover:  
$M^{new}_{i\tilde{A}_{new}}$ and $N^{new}_{i\tilde{\tilde{B}}_{new}}$.  
A concrete example of crossover of simplicial complex matrices corresponding to the example in Figure \ref{graph_crossover} is shown in Figure \ref{matrix_crossover}.

Appendix B.2 provides the implementation of the genetic algorithm that performs mutation and crossover operations on simplicial complex matrices.

\section{Conclusion and discussion}
Colored graphs and their subgraphs, called bubble graphs, correspond one-to-one
to simplicial complexes of pseudo-manifolds and their subsimplices.
In this paper, we introduced a matrix representation of colored graphs and their
bubble graphs, and, using this correspondence, defined simplicial complex
matrices and subsimplex matrices that represent simplicial complexes of
pseudo-manifolds and their subsimplices.

Furthermore, we defined mutations and crossovers of colored graphs, and, by
using the correspondence among simplicial complexes, colored graphs, and
simplicial complex matrices, we defined mutations and crossovers of simplicial
complexes and simplicial complex matrices.

We also implemented code that generates simplicial complex matrices and performs
their mutations and crossovers, thereby realizing a genetic algorithm capable
of generating pseudo-manifolds with various topologies.
In addition, we created code that decomposes the generated simplicial complex
matrix into simplex matrices, reconstructs the simplicial complex of the
pseudo-manifold from this information, and computes the volume of each simplex,
its circumcenter, and the volume of its dual simplex.

The next step is to clarify the geometric meaning of the mutations and
crossovers of pseudo-manifolds defined via this correspondence.
Another possible direction is that, since various topologies can now be
generated numerically, we can apply this framework to search for the minimum of
discretized potential for string backgrounds using Regge calculus \cite{Hatakeyama:2024fzv, future}.
This may allow us to determine the vacuum of string theory, and from
fluctuations around it, make predictions for corrections to the Standard Model, inflationary cosmology, and so on.

\section{Acknowledgement}
We would like to thank
D. Kadoh,
I. Kanamori,
H. Kawai,
T. Masuda,
K. Nagasaki,
J. Nishimura,
T. Onogi,
Y. Sugimoto,
M. Takeuchi,
T. Yoneya,
and especially
A. Tsuchiya
for discussions.

\appendix

\section{Dual polytopes of simplices and their volumes}
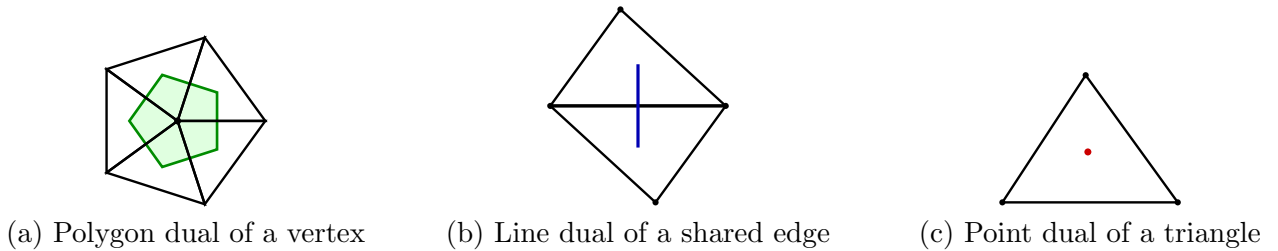
\begin{figure}[h]
    \centering
    \begin{minipage}[t]{0.32\linewidth}
        \centering
        \begin{tikzpicture}[scale=0.58, transform shape]
            \coordinate (v)  at (0,0);
            \coordinate (a1) at (2.0,0.0);
            \coordinate (a2) at (0.62,1.90);
            \coordinate (a3) at (-1.62,1.18);
            \coordinate (a4) at (-1.62,-1.18);
            \coordinate (a5) at (0.62,-1.90);

            \coordinate (c1) at (0.90,0.65);
            \coordinate (c2) at (-0.35,1.05);
            \coordinate (c3) at (-1.10,0.00);
            \coordinate (c4) at (-0.35,-1.05);
            \coordinate (c5) at (0.90,-0.65);

            \filldraw[fill=green!12, draw=green!55!black, line width=1.0pt]
                (c1) -- (c2) -- (c3) -- (c4) -- (c5) -- cycle;

            \draw[line width=0.9pt] (v) -- (a1) -- (a2) -- cycle;
            \draw[line width=0.9pt] (v) -- (a2) -- (a3) -- cycle;
            \draw[line width=0.9pt] (v) -- (a3) -- (a4) -- cycle;
            \draw[line width=0.9pt] (v) -- (a4) -- (a5) -- cycle;
            \draw[line width=0.9pt] (v) -- (a5) -- (a1) -- cycle;

            \fill[black] (v) circle [radius=0.08];
        \end{tikzpicture}

        \small (a) Polygon dual of a vertex
    \end{minipage}
    \hfill
    \begin{minipage}[t]{0.32\linewidth}
        \centering
        \begin{tikzpicture}[scale=0.58, transform shape]
            \coordinate (A) at (0,0);
            \coordinate (B) at (4,0);
            \coordinate (C) at (1.6,2.2);
            \coordinate (D) at (2.4,-2.2);

            \coordinate (u) at (2.0,0.95);
            \coordinate (w) at (2.0,-0.95);

            \draw[line width=0.9pt] (A) -- (B) -- (C) -- cycle;
            \draw[line width=0.9pt] (A) -- (B) -- (D) -- cycle;

            \draw[very thick, black] (A) -- (B);
            \draw[blue!70!black, line width=1.2pt] (u) -- (w);

            \fill[black] (A) circle [radius=0.07];
            \fill[black] (B) circle [radius=0.07];
            \fill[black] (C) circle [radius=0.07];
            \fill[black] (D) circle [radius=0.07];
        \end{tikzpicture}

        \small (b) Line dual of a shared edge
    \end{minipage}
    \hfill
    \begin{minipage}[t]{0.32\linewidth}
        \centering
        \begin{tikzpicture}[scale=0.58, transform shape]
            \coordinate (p1) at (0,0);
            \coordinate (p2) at (4,0);
            \coordinate (p3) at (1.9,2.9);
            \coordinate (ct) at (1.95,1.15);

            \draw[line width=0.9pt] (p1) -- (p2) -- (p3) -- cycle;
            \fill[black] (p1) circle [radius=0.07];
            \fill[black] (p2) circle [radius=0.07];
            \fill[black] (p3) circle [radius=0.07];
            \fill[red!80!black] (ct) circle [radius=0.08];
        \end{tikzpicture}

        \small (c) Point dual of a triangle
    \end{minipage}
    \caption{Dual objects in two dimensions shown in one figure: (a) the dual of a vertex with five incident triangles is a polygon, (b) the dual of an edge shared by two triangles is a line segment, and (c) the dual of a triangle is a point.}
    \label{dual_objects}
\end{figure}

On a simplicial complex, we can construct dual polytope ${}^*T^p$ for each $p$-simplex $T^p$.
The definition is as follows: To any given $p$-simplex $T^p$,
we associate the flat $(D-p)$-dimensional polytope ${}^*T^p$ that consists of all points which are at equal distances of all the vertices of $T^p$, but closer to these vertices than to the other vertices of the complex.
Examples in the two dimensions are depicted in Figure~\ref{dual_objects}.

A recursive expression for the algebraic volume $V({}^*T^p)$ of the dual polytope ${}^*T^p$ is given by 
\cite{Ren:1987is},
\begin{align}
	V({}^*T^p)
	=& \frac{1}{D-p} \sum_{T^{p+1} \ni T^p} V({}^*T^{p+1})\, \chi(T^{p+1}, T^p),
\end{align}
where the sum is taken over all $(p+1)$-simplices $T^{p+1}$ that contain $T^p$ as a face, and $\chi(T^{p+1}, T^p)$ is a signed distance from the circumcenter of $T^{p+1}$ to $T^p$.
The sign is determined as follows: Consider the $(p+1)$-dimensional hyperspace containing $T^{p+1}$ and hyperplane in that hyperspace containing $T^p$, which divides the space into two half-spaces. If the circumcenter of $T^{p+1}$ is in the same half-space as the vertex of $T^{p+1}$ opposite to $T^p$, let $\chi(T^{p+1}, T^p) > 0$, otherwise, $\chi(T^{p+1}, T^p) < 0$.
We set $V({}^*T^D) = 1$ for all $D$-simplices $T^D$.
Once the simplicial complex structure and the starting simplex $T^p$ are given,
the detail of recursive summation is completely determined.

\begin{figure}[h]
  \centering
  \includegraphics[width=100mm]{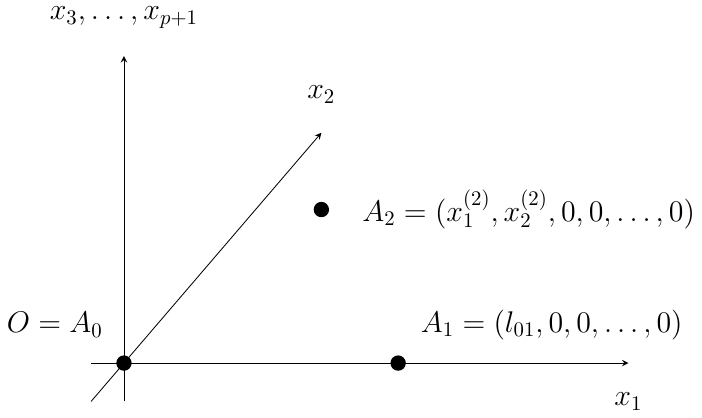}
  \caption{the coordinates.}
  \label{decartesian}
\end{figure}

To determine the sign of the signed distance $\chi(T^{p+1}, T^p)$, we express
the coordinates of the circumcenter $G_{p+1}$ and the vertices $A_0$, $A_1$, $\dots$, $A_p$,  $A_{p+1}$ of a $(p+1)$-simplex
$T^{p+1}$ in terms of edge
lengths, and use this information to determine the position of $G_{p+1}$.
Let $l_{ij}$ denote the length of the edge between $A_i$ and $A_j$, all assumed
to be known.
For this purpose, we use the following lemma:
\begin{lem}
The circumcenter $G_{p+1}$ of a $(p+1)$-simplex $T^{p+1}$ lies on the line passing through the
circumcenter $G_p$ of any of its $p$-subsimplices, and this line is orthogonal
to that $p$-subsimplex.
\label{circumcentre}
\end{lem}
First, we define an orthogonal coordinate system consisting of the $p+1$ directions
$x_1, \allowbreak x_2, \allowbreak x_3, \allowbreak \dots, \allowbreak x_{p+1}$,
and express the coordinates of all vertices of $T^{p+1}$ in this coordinate system
in terms of the edge lengths of $T^{p+1}$.
Let the coordinates of the vertex $A_i$ in this coordinate system be
$(x^{(i)}_1, x^{(i)}_2, \dots, x^{(i)}_{p+1})$.
This coordinate system is defined as follows (Fig.~\ref{decartesian}).
The origin $O$ is taken to coincide with the vertex $A_0$.
Each axis of the coordinate system is chosen so that
the coordinates of the vertex $A_1$ become $(l_{01},0,0,\dots,0)$,
and furthermore, for each vertex $A_i$, its $i$-th coordinate component is positive
and all components from the $(i+1)$-th onward are zero.
Under this definition, the $(i+1)$-simplex consisting of the vertices
$A_0, A_1, A_2, A_3, \dots, A_i$ lies in the space spanned by the axes
$x_1, x_2, x_3, \dots, x_i$ of this coordinate system.

Under this choice of coordinates,  
suppose that the coordinates of the vertices $A_1, A_2, \dots, A_k$  
and the coordinates of the circumcenter $G_k$ of the $k$-face $T^k$  
formed by the vertices $A_0, A_1, A_2, \dots, A_k$
are already expressed in terms of the edge lengths of $T^k$.
Now consider the $(k+1)$-face $T^{k+1}$ formed by the vertices
$A_0, A_1, A_2, \dots, A_{k+1}$.
The coordinates of the vertex $A_{k+1}$ can be expressed in terms of
the coordinates of $A_1, A_2, \dots, A_k$ and the distances from
$A_0, A_1, A_2, \dots, A_k$ to $A_{k+1}$, i.e., using the edge lengths of $T^{k+1}$.
Strictly speaking, these conditions do not determine the sign of the
$(k+1)$-th component $x^{(k+1)}_{k+1}$ of $A_{k+1}$,
but our coordinate system is defined so that $x^{(k+1)}_{k+1}$ is positive.

Now, if the coordinates of $G_k$ are written as
$G_k = (g^{(k)}_1, g^{(k)}_2, \dots, g^{(k)}_k, 0, \dots, 0)$,
then the coordinates of the circumcenter $G_{k+1}$ of $T^{k+1}$ are
\[
G_{k+1} = (g^{(k)}_1, g^{(k)}_2, \dots, g^{(k)}_k, g^{(k+1)}_{k+1}, 0, \dots, 0).
\]
Here we used Lemma~\ref{circumcentre}.
Since the $g^{(k)}_i$ are already known, determining $g^{(k+1)}_{k+1}$
gives the coordinates of $G_{k+1}$.
Because $x^{(p+1)}_{p+1}$ of $A_{p+1}$ is defined to be positive, 
$A_{p+1}$ and the circumcenter are in the same side of $T^p$ if $g^{(p+1)}_{p+1}$ is positive, and they are in the other side otherwise. Thus, the sign of $\chi (T^{p+1}, T^p)$ is determined and $\chi (T^{p+1}, T^p) = g^{(p+1)}_{p+1}$. 

On the other hand, the distance $R_{k+1}$ from $G_{k+1}$ to each vertex of $T^{k+1}$
(the circumradius) can be computed using the following formula~\cite{Coxeter}:
\begin{align}
	R_{k+1}^2 &= - \frac{1}{2} \frac{\det U_{k+1}}{\det W_{k+1}},	\nonumber	\\
	U_{k+1} &=
		\begin{pmatrix}
			0	&	l_{01}^2	&	l_{02}^2	& \dots	& l_{0,k+1}^2	\\
			l_{10}^2	&	0	&	l_{12}^2	& \dots	& l_{1,k+1}^2	\\
			l_{20}^2	&	l_{21}^2	&	0	& \dots	& l_{2,k+1}^2	\\
			\vdots	&	\vdots	&	\vdots	&	\cdots	&	\vdots	\\
			l_{k+1,0}^2	&	l_{k+1,1}^2	&	l_{k+1,2}^2	& \dots	& 0
		\end{pmatrix},	\nonumber	\\
	W_{k+1} &=
		\begin{pmatrix}
			0	&	l_{01}^2	&	l_{02}^2	& \dots	& l_{0,k+1}^2	&	1	\\
			l_{10}^2	&	0	&	l_{12}^2	& \dots	& l_{1,k+1}^2	&	1	\\
			l_{20}^2	&	l_{21}^2	&	0	& \dots	& l_{2,k+1}^2	&	1	\\
			\vdots	&	\vdots	&	\vdots	&	\cdots	&	\vdots	&	\vdots	\\
			l_{k+1,0}^2	&	l_{k+1,1}^2	&	l_{k+1,2}^2	& \dots	& 0	&	1	\\
			1	&	1	&	1	&	\cdots	&	1	&	0
		\end{pmatrix}.
\end{align}
Using the value of $R_{k+1}$ obtained from this formula, the volume
$V_{G,k+1}$ of the $(k+1)$-simplex $\{ A_0, A_1, \dots, A_k, G_{k+1} \}$
can be computed by applying the Cayley-Menger determinant
formula~\cite{liberti2015mathematical}:
\begin{align}
	V_{G,k+1}^2 =  \frac{(-1)^{k+2}}{(k+1)!(k+1)!2^{k+1}}
		\begin{vmatrix}
			0	&	l_{01}^2	&	l_{02}^2	& \dots	& l_{0,G_{k+1}}^2	&	1	\\
			l_{10}^2	&	0	&	l_{12}^2	& \dots	& l_{1,G_{k+1}}^2	&	1	\\
			l_{20}^2	&	l_{21}^2	&	0	& \dots	& l_{2,G_{k+1}}^2	&	1	\\
			\vdots	&	\vdots	&	\vdots	&	\cdots	&	\vdots	&	\vdots	\\
			l_{0,G_{k+1}}^2	&	l_{1,G_{k+1}}^2	&	l_{2,G_{k+1}}^2	& \dots	& 0	&	1	\\
			1	&	1	&	1	&	\cdots	&	1	&	0
		\end{vmatrix}.
\end{align}
Here, $l_{i,G_{k+1}} = R_{k+1}$.

On the other hand, $V_{G,k+1}$ can be written in terms of the volume $V_k$
of the $k$-simplex $\{A_0, \allowbreak A_1, \allowbreak \dots, \allowbreak A_k\}$ as
\begin{align}
	V_{G,k+1} = \frac{1}{k+1} V_k |g^{(k+1)}_{k+1}|.
\end{align}
Here again we have used Lemma~\ref{circumcentre}.
Combining these results gives
\begin{align}
	|g^{(k+1)}_{k+1}| = \frac{V_{G,k+1}}{V_k} (k+1),
\end{align}
which is expressible in terms of the edge lengths of $T^{k+1}$.
Thus,
\begin{align}
	G_{k+1} = \left(g^{(k)}_1, g^{(k)}_2, \dots, g^{(k)}_k,
		\pm \frac{V_{G,k+1}}{V_k} (k+1), 0, \dots, 0 \right)
\end{align}
follows.
Finally, the sign of the $(k+1)$-th component of $G_{k+1}$ is determined
from the condition that the distance between $G_{k+1}$ and $A_{k+1}$
equals $R_{k+1}$.

So far, we have assumed that the coordinates of vertices
$A_1, A_2, \dots, A_k$ and the coordinates of the circumcenter $G_k$
of the $k$-face $T^k$ consisting of $A_0, A_1, A_2, \dots, A_k$
are known in terms of edge lengths.
We showed that under this assumption, the coordinates of the vertices
$A_1, A_2, \dots, A_{k+1}$ and the coordinates of the circumcenter $G_{k+1}$
of the $(k+1)$-face $T^{k+1}$ consisting of
$A_0, \allowbreak A_1, \allowbreak A_2, \allowbreak \dots, \allowbreak A_{k+1}$ can also be expressed in terms of edge lengths.
Carrying this out inductively from $k=1$, we can finally determine
the coordinates of $A_0, A_1, \dots, A_{p+1}$ and of the circumcenter $G_{p+1}$
of the $(p+1)$-simplex.

Appendix B.6 contains code which, given a simplicial complex,
computes for each $p$-simplex its circumradius, circumcenter,
the signed distance $\chi$ from the circumcenter to its base face,
and the volume $V_G$ of the simplex obtained by replacing one vertex
with the circumcenter.
Appendix B.7 contains code that computes the algebraic volume $V({}^*T^p)$ of the dual polytope
${}^*T^p$ corresponding to each $p$-simplex $T^p$ of the simplicial complex.

\section{Codes for numerical simulations}
This appendix provides an overview of how to use the numerical simulation codes developed in this study, located in \cite{codes}.

\subsection{Generation of simplicial complex matrices}
The program file \path|geom_generate_mat.py| is used to generate the initial simplicial complex matrices. When executed, it produces data files \path|initial_generation_2D.bin| or \path|initial_generation_6D.bin|.

By modifying the section below, the user can adjust the output.
\begin{verbatim}
### main ###
edgecolor = ['red','orange','black','green','blue','gray','purple']
#edgecolor = ['red','blue','black']
\end{verbatim}
The list \verb|edgecolor| specifies the set of colors, where selecting three colors corresponds to the two-dimensional case, and selecting seven colors corresponds to the six-dimensional case.

\subsection{Genetic algorithm for simplicial complex matrices}
The program file \verb|geom_v4.py| generates new simplicial complex matrices in accordance with a genetic algorithm. Upon execution, this program reads \verb|initial_generation_2D.bin| or \linebreak \verb|initial_generation_6D.bin| and generates a folder, \verb|results_2D| or \verb|results_6D| that includes matrix files,  \verb|gen_xxx_sizexxx.dat|, where the entries are ordered sequentially column by column.

The following parameters can be adjusted to tune the behavior of the genetic algorithm. 
\begin{verbatim}
### main ###
#edgecolor = ['red','blue','black']
edgecolor = ['red','orange','black','green','blue','gray','purple']

P_Crossover = 0.2
P_Mutation =  0.5
Population = 300
N_loop = 10
Tournament_size = 3
\end{verbatim}
The parameter \verb|edgecolor| functions identically to that described in the preceding subsection. The variables \verb|P_Crossover| and \verb|P_Mutation| denote the crossover and mutation probabilities, respectively. The variable \verb|Population| specifies the number of simplicial complex matrices, while \verb|N_loop| determines the number of iterations of crossover and mutation operations. The selection procedure chooses matrices for the next generation by preferentially retaining those of larger size; specifically, \verb|Tournament_size| matrices are sampled from the population, and selection is performed among them.

\subsection{Decomposition of a simplicial complex matrix into simplex matrices}
The program file \verb|generate_simplex_v10.f90| decomposes a given simplicial complex matrix generated by the Python code into matrices representing each $p$-simplex. The input file is \linebreak \verb|in_generate_for_n_simplex.dat|, which has the following structure:

\begin{verbatim}
3                                   !the number of rows
'gen_xxx_sizexxx.dat'               !input file
'out_simplex_0.dat'                 !output file 0
'out_simplex_1.dat'                 !output file 1
\end{verbatim}

\begin{equation}
 \vdots \qquad\qquad\qquad\qquad\qquad\qquad\qquad\qquad\qquad\qquad\qquad\qquad
 \nonumber
\end{equation}
Here, the first line specifies the number of rows of the matrix. The second line gives the name of the input matrix file generated by the Python code.
The lines following the second specify the output filenames for each value of $p$.

This code also generates \verb|in_identify_simplex_v9.dat|, which will be used in the second code.

\subsection{Construction of the simplicial complex}
\begin{figure}[htb]
\centering
\includegraphics[width=15cm]{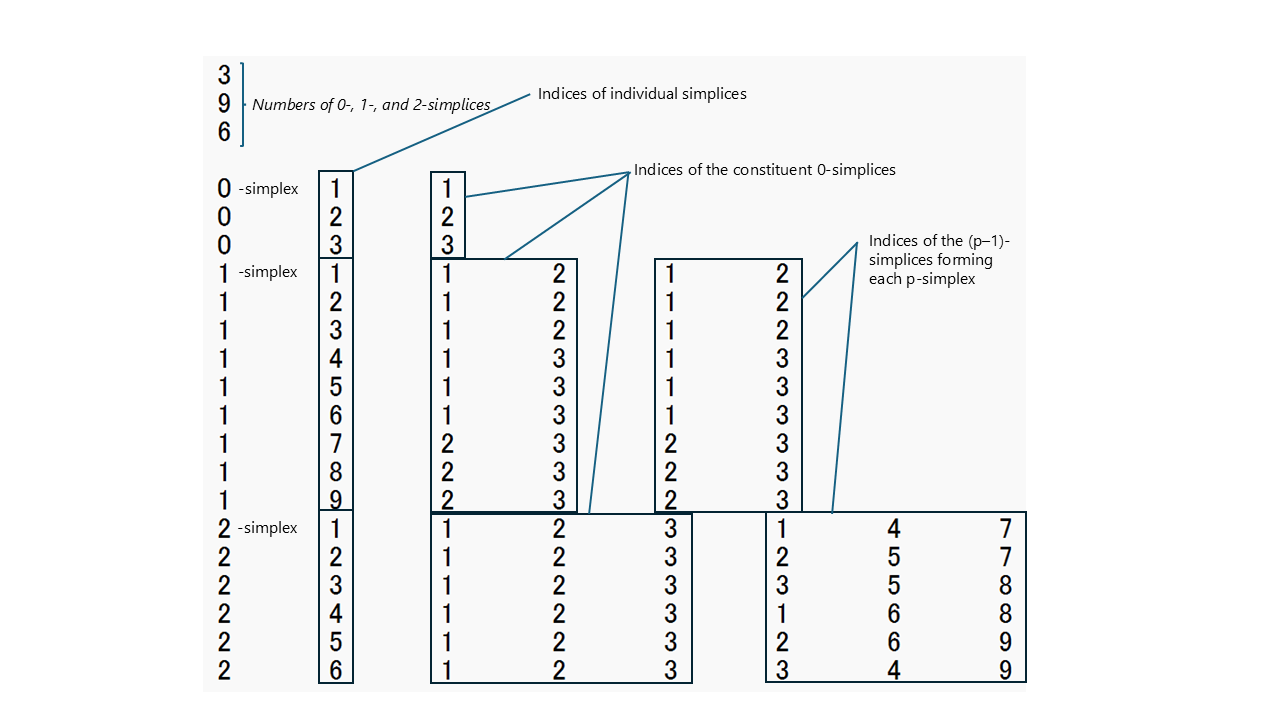}
\caption{Structure of the simplicial complex}
\label{OutIdentify}
\end{figure}

The program file \verb|identify_simplex_v9.f90| generates the geometric data of a simplicial complex based on the simplex matrices produced by \verb|generate_simplex_v7.f90| in the previous subsection. The input file is \verb|in_identify_simplex_v9.dat|, which has the following structure:
\begin{verbatim}
2                       !dim of manifold, dim = the number of rows-1
6                       !the number of columns
'out_identify_2dim.dat' !output file
'out_simplex_0.dat'     !input file 0
'out_simplex_1.dat'     !input file 1
'out_simplex_2.dat'     !input file 2
\end{verbatim}
Here, the first line specifies the dimension of the simplicial complex, which is equal to the number of matrix rows minus one; the second line gives the number of columns; the third line specifies the output file name (\verb|out_identify_2dim.dat|); and the subsequent lines list the output files for each $p$ from \verb|generate_simplex_v7.f90|. The file \verb|out_identify_2dim.dat| has the structure illustrated in Figure~\ref{OutIdentify}.

In $D$ dimensions, the output file \path|out_simplex_D.dat| generated by \path|generate_simplex_v7.f90| is empty. This is because the corresponding matrix becomes the zero matrix once all $D+1$ rows are removed. In contrast, the output file \path|out_identify_Ddim.dat| produced by \path|identify_simplex_v9.f90| contains information about the $D$-simplices. A $D$-simplex corresponds to a single vertex, and then
contains a $p$-simplex 
whose column associated to that vertex  have nonzero entries. This follows from the fact that the graph representing each $p$-simplex contains the vertex corresponding to the nonzero column and connects it to vertices corresponding to the other columns involved.

This code also generates  \verb|in_length_volume_v6.dat|, which will be used in the second code. 

\subsection{Volumes of simplices}
The program file \verb|length_volume_v6.f90| determines the volumes of $p$-simplices by first assigning random lengths to all 1-simplices. Although these lengths are randomly generated, three lines of code are provided for overwriting all edge lengths with unity for verification purposes. These lines can yield the standard simplex volumes, although they are commented out as default. The input file is \verb|in_length_volume_v6.dat|, which has the following structure:
\begin{verbatim}
2                             !dim of manifold, dim = N_row-1
'out_identify_2dim.dat'       !input file
'out_length_volume_2dim.dat'  !output file
\end{verbatim}
The first line specifies the dimension, the second line the output file from \path|identify_simplex_v9.f90|, and the third line the name of the output file.

This code also generates  \verb|in_circumcenter_v9.dat|, which will be used in the second code. 

\subsection{Circumcenters of simplices}
The program file \verb|circumcenter_v9.f90| computes for each $p$-simplex its circumradius, circumcenter, the signed distance $\chi$ from the circumcenter to the base face, and the volume $V_G$ of the simplex obtained by replacing one of its vertices with the circumcenter. During compilation, the LAPACK and BLAS linear algebra libraries must be linked using the flags \verb|-llapack -lblas|. The input file is \verb|in_circumcenter_v9.dat|, which has the following structure:
\begin{verbatim}
2                             !dim of manifold
'out_identify_2dim.dat'       !input file
'out_length_volume_2dim.dat'  !input file of volume for p-simplex
'out_circumcenter_2dim.dat'   !output file
\end{verbatim}
Here, the first line gives the dimension; the second line gives the output file of \path|identify_simplex_v9.f90|; the third line that of \path|length_volume_v6.f90|; and the fourth line specifies the output file.

This code also generates  \verb|in_dual_volume_v4.dat|, which will be used in the second code. 

\subsection{Volumes of dual polytopes}
The program file \verb|dual_volume_v4.f90| computes for each $p$-simplex $T^p$ the algebraic volume of its dual polytope $*T^p$. Compilation requires linking the LAPACK and BLAS libraries via \verb|-llapack -lblas|. The input file is \verb|in_dual_volume_v4.dat|, which has the following structure:
\begin{verbatim}
2                              !dim of manifold
'out_identify_2dim.dat'        !input file
'out_length_volume_2dim.dat'   !input file of volume for p-simplex
'out_circumcenter_2dim.dat'    !input file of radius and position of circumcenter
'out_dual_volume_2dim.dat'     !output file
\end{verbatim}
The first line specifies the dimension.  
The second, third, and fourth lines give the output files of \verb|identify_simplex_v9.f90|, \verb|length_volume_v6.f90|, and \verb|circumcenter_v9.f90|, respectively.  
The fifth line provides the name of the output file, which contains the simplex volume, the dual volume, $V_G$, and $\chi$ for each $p$-simplex.


\begin{thebibliography}{100}

\bibitem{Sato:2017qhj}
M.~Sato, ``{String geometry and nonperturbative formulation of string
  theory},'' \href{http://dx.doi.org/10.1142/S0217751X19501264}{ Int. J.
  Mod. Phys. A {\bfseries 34} no.~23, (2019) 23},
  \href{http://arxiv.org/abs/1709.03506}{{\ttfamily arXiv:1709.03506
  [hep-th]}}.
  

\bibitem{Sato:2019cno}
M.~Sato and Y.~Sugimoto, ``{Topological string geometry},''
  \href{http://dx.doi.org/10.1016/j.nuclphysb.2020.115019}{ Nucl. Phys. B
  {\bfseries 956} (2020) 115019},
  \href{http://arxiv.org/abs/1903.05775}{{\ttfamily arXiv:1903.05775
  [hep-th]}}.

\bibitem{Sato:2020szq}
M.~Sato and Y.~Sugimoto, ``{Perturbative string theory from Newtonian limit of
  string geometry theory},''
  \href{http://dx.doi.org/10.1140/epjc/s10052-020-8255-5}{ Eur. Phys. J. C
  {\bfseries 80} no.~8, (2020) 789},
  \href{http://arxiv.org/abs/2002.01774}{{\ttfamily arXiv:2002.01774
  [hep-th]}}.

\bibitem{Sato:2022owj}
M.~Sato, Y.~Sugimoto, and K.~Uzawa, ``{Path integrals of perturbative strings
  on curved backgrounds from string geometry theory},''
  \href{http://dx.doi.org/10.1103/PhysRevD.106.086006}{ Phys. Rev. D
  {\bfseries 106} no.~8, (2022) 086006},
  \href{http://arxiv.org/abs/2203.16304}{{\ttfamily arXiv:2203.16304
  [hep-th]}}.

\bibitem{Sato:2022brv}
M.~Sato and K.~Uzawa, ``{Path integrals of perturbative superstrings on curved
  backgrounds from string geometry theory},''
  \href{http://dx.doi.org/10.1103/PhysRevD.107.066023}{ Phys. Rev. D
  {\bfseries 107} no.~6, (2023) 066023},
  \href{http://arxiv.org/abs/2211.16959}{{\ttfamily arXiv:2211.16959
  [hep-th]}}.

  

\bibitem{Sato:2023lls}
M.~Sato and T.~Tohshima,
``T-symmetry in String Geometry Theory,''
Adv. High Energy Phys. \textbf{2025}, 7148232 (2025)
[arXiv:2301.08952 [hep-th]].


 \bibitem{Honda:2020sbl}
M.~Honda and M.~Sato, ``{String Backgrounds in String Geometry},''
  \href{http://dx.doi.org/10.1142/S0217751X20501766}{ Int. J. Mod. Phys. A
  {\bfseries 35} no.~27, (2020) 27},
  \href{http://arxiv.org/abs/2003.12487}{{\ttfamily arXiv:2003.12487
  [hep-th]}}.

\bibitem{Honda:2021rcd}
M.~Honda, M.~Sato, and T.~Tohshima, ``{Superstring Backgrounds in String
  Geometry},'' \href{http://dx.doi.org/10.1155/2021/9993903}{ Adv. High
  Energy Phys. {\bfseries 2021} (2021) 9993903},
  \href{http://arxiv.org/abs/2102.12779}{{\ttfamily arXiv:2102.12779
  [hep-th]}}. 


  
  
  

\bibitem{Nagasaki:2023fnz}
K.~Nagasaki, M.~Sato and G.~Tanaka,
``The perturbative vacua in string geometry theory,''
[arXiv:2309.10394 [hep-th]].


\bibitem{Sato:2025wfc}
M.~Sato,
``Fundamental structure of string geometry theory,''
[arXiv:2511.02310 [hep-th]].



\bibitem{Nagasaki:2025tmi}
K.~Nagasaki and M.~Sato,
``The heterotic perturbative vacua in string geometry theory,''
[arXiv:2511.03357 [hep-th]].



\bibitem{Sato:2025qqa}
M.~Sato and M.~Takeuchi,
``String geometry phenomenology,''
[arXiv:2511.04145 [hep-th]].



\bibitem{futureIIB}
R. Kudo, K. Nagasaki and M.~Sato, in preparation.





\bibitem{Regge:1961px}
T.~Regge,
``GENERAL RELATIVITY WITHOUT COORDINATES,''
Nuovo Cim. \textbf{19}, 558-571 (1961)

\bibitem{Ambjorn:1998xu}
J.~Ambjorn and R.~Loll,
``Nonperturbative Lorentzian quantum gravity, causality and topology change,''
Nucl. Phys. B \textbf{536}, 407-434 (1998)
[arXiv:hep-th/9805108 [hep-th]].

\bibitem{CHRIST1982337}
N.~H.~Christ, R.~Friedberg, and T.~D.~Lee,
``Weights of links and plaquettes in a random lattice,''
Nucl. Phys. B \textbf{210}, 337-346 (1982)


\bibitem{Ren:1987is}
H.~c.~Ren,
``Matter Fields in Lattice Gravity,''
Nucl. Phys. B \textbf{301}, 661-684 (1988)
doi:10.1016/0550-3213(88)90281-7


\bibitem{Gurau:2011xp}
R.~Gurau and J.~P.~Ryan,
``Colored Tensor Models - a review,''
SIGMA \textbf{8}, 020 (2012)
doi:10.3842/SIGMA.2012.020
[arXiv:1109.4812 [hep-th]].


\bibitem{Hatakeyama:2024fzv}
K.~Hatakeyama, M.~Sato and G.~Tanaka,
[arXiv:2404.19413 [hep-lat]].

 \bibitem{future}
K. Hatakeyama, K. Nagasaki, M.~Sato, Y.~Sugimoto, and G. Tanaka, work in progress.

\bibitem{David:1992jw}
F.~David,
``Simplicial quantum gravity and random lattices,''
[arXiv:hep-th/9303127 [hep-th]].



 \bibitem{Coxeter}
H.~S.~M.~Coxeter, ``The circumradius of the general simplex,'' The Mathematical Gazette, 15(210):229-231, 1930.

 \bibitem{liberti2015mathematical}
 L.~Liberti and C.~Lavor, ``Six mathematical gems from the history of distance geometry,'' arXiv:1502.02816 [math.HO].
 
 \bibitem{codes}
https://doi.org/10.5281/zenodo.20743401



\end{thebibliography}
\end{document}